\documentclass[
 reprint,
superscriptaddress,
amsmath,amssymb,
aps,
pra,
]{revtex4-2}

\usepackage{graphicx}
\usepackage{dcolumn}
\usepackage{bm}
\usepackage{gensymb}
\usepackage{url}
\usepackage{multirow}

\usepackage{xcolor}

\begin{document}

\preprint{APS/123-QED}

\title{Shaping the quantum vacuum with anisotropic temporal boundaries}

\author{J. Enrique V\'azquez-Lozano}
\affiliation{%
 Department of Electrical, Electronic and Communications Engineering, Institute of Smart Cities (ISC), Public University of Navarre (UPNA), 31006 Pamplona, Spain
}%

\author{Iñigo Liberal}
\thanks{Corresponding author: inigo.liberal@unavarra.es}%
\affiliation{%
 Department of Electrical, Electronic and Communications Engineering, Institute of Smart Cities (ISC), Public University of Navarre (UPNA), 31006 Pamplona, Spain
}%

\begin{abstract}
Temporal metamaterials empower novel forms of wave manipulation with direct applications to quantum state transformations. In this work, we investigate vacuum amplification effects in anisotropic temporal boundaries. Our results theoretically demonstrate that the anisotropy of the temporal boundary provides control over the angular distribution of the generated photons. We analyze several single and multi-layered configurations of anisotropic temporal boundaries, each with a distinct vacuum amplification effect. Examples include the inhibition of photon production along specific directions, resonant and directive vacuum amplification, the generation of angular and frequency photon combs, and fast angular variations between inhibition and resonant photon production.
\end{abstract}

\maketitle

\section{Introduction}

Temporal metamaterials or time-varying media offer advanced wave manipulation opportunities via simultaneous control of spatial and temporal degrees of freedom \cite{Caloz2019spacetime,galiffi2022photonics,Yin2022}. The basic process underpinning temporal metamaterials is that of a temporal boundary/time interface, i.e., an abrupt (theoretically instantaneous) change in material properties at a localized point in time. Early works \cite{Morgenthaler1958velocity,Felsen1970wave,Fante1971transmission,Fante1973propagation,xiao2014reflection} identified that temporal boundaries result in mixing of forward and backward waves, shifted in frequency with respect to the waves before the temporal boundary. However, recent research on time-varying media has explored the many opportunities offered by more complex metamaterial temporal boundaries.

For example, anisotropic temporal boundaries act as an ``inverse prism" \cite{akbarzadeh2018inverse}, can redirect the energy of propagating waves \cite{Pacheco2020temporal}, and exhibit temporal Brewster angles where no backward wave is produced \cite{pacheco2021Brewster}. Frequency dispersive temporal boundaries enable multi-frequency generation \cite{Solis2021time,Gratus2021temporal}, and nonreciprocal temporal boundaries exhibit Faraday rotation effects \cite{Li2022nonreciprocity}. 
Combining two or more boundaries into a temporally multi-layered system provides further design flexibility, including control over the backward wave and its spectral response \cite{Pacheco2020antireflection,Castaldi2021exploiting,Ramaccia2020light,Ramaccia2021temporal}.
Furthermore, when a large number of temporal boundaries are put together, the system can be effectively described as a photonic time crystal \cite{Lustig2018topological,Sharabi2021disordered,Sharabi2022spatiotemporal} or a space-time metamaterial \cite{Huidobro2021homogenization} granting access to new forms of light propagation. 

\begin{figure*}[!t]
\includegraphics[width=6in]{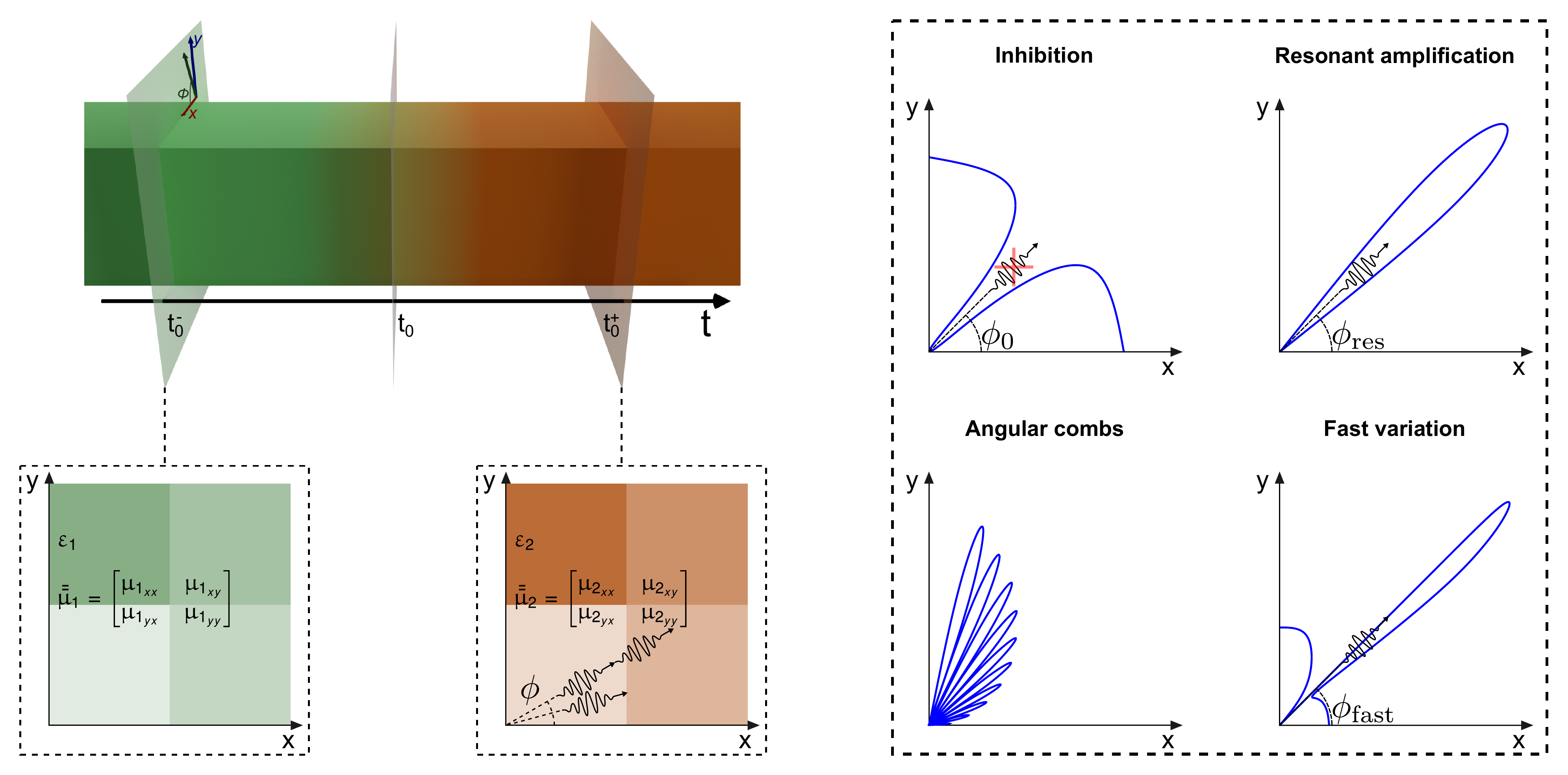}
\caption{{\bf Schematic depiction of an anisotropic temporal boundary.} At time $t_0$, the material parameters of a two-dimensional (2D) system suddenly change from $(\boldsymbol{\mu}_1,\varepsilon_{z1})$ to $(\boldsymbol{\mu}_2,\varepsilon_{z2})$. The interaction of vacuum fluctuations with the dynamical boundary result in photon production with a nontrivial angular profile. Engineering the anisotropy of the boundary and its time sequencing  enables a variety of vacuum amplification effects, including inhibition of photon production along specific directions, resonant and directive amplification, the generation of angular combs, and fast angular variations from inhibition to resonant amplification.}
\label{fig:Fig1}
\end{figure*}

Temporal boundaries are also of interest for the field of quantum optics, where they have been shown to result in squeezing transformations \cite{Mendoncca2000quantum,Mendoncca2003temporal,Mendoncca2005time}. They also modify light emission from quantum emitters \cite{Lyubarov2022amplified} and free electrons \cite{Dikopoltsev2022light}.
Similar to the classical case, it is expected that the design flexibility offered by metamaterials will open new pathways for research on quantum time-varying media. Following this motivation, in this work we investigate how anisotropic temporal boundaries provide control over the angular properties of vacuum amplification effects (see Fig.\,\ref{fig:Fig1}). Vacuum amplifications effects \cite{Nation2012colloquium,Dodonov2020fifty} consist of photon generation from the electromagnetic vacuum state, produced by the interaction between quantum vacuum fluctuations and a dynamic boundary. As schematically depicted in Fig.\,\ref{fig:Fig1}, anisotropic temporal boundaries allow for controlling the angular distribution of the generated photons, including inhibiting the production of photons along a specific direction, resonantly enhancing photon emission while concentrating all of them into a single direction, generating frequency and angular combs of photons, and producing fast angular variations from zero to resonant amplification, akin to Fano resonances. 

Most experiments on vacuum amplification effects have been carried out at microwave frequencies by using superconducting circuits \cite{Wilson2011observation,Lahteenmaki2013dynamical}. Such technological platform resonates with experiments on transmission line metamaterials (TL-MTMs) \cite{Caloz2005,Eleftheriades2005negative}, which enabled the first observation of metamaterial lenses overcoming the diffraction limit \cite{Grbic2004overcoming}. The extension of TL-MTMs to anisotropic systems has also been experimentally demonstrated as an enabler of transformation optics applications \cite{Zedler2011anisotropic,Wong2006fields}.

\section{Anisotropic temporal boundaries}

\subsection{Theory of anisotropic temporal boundaries}

\begin{figure}[!t]
\includegraphics[width=3in]{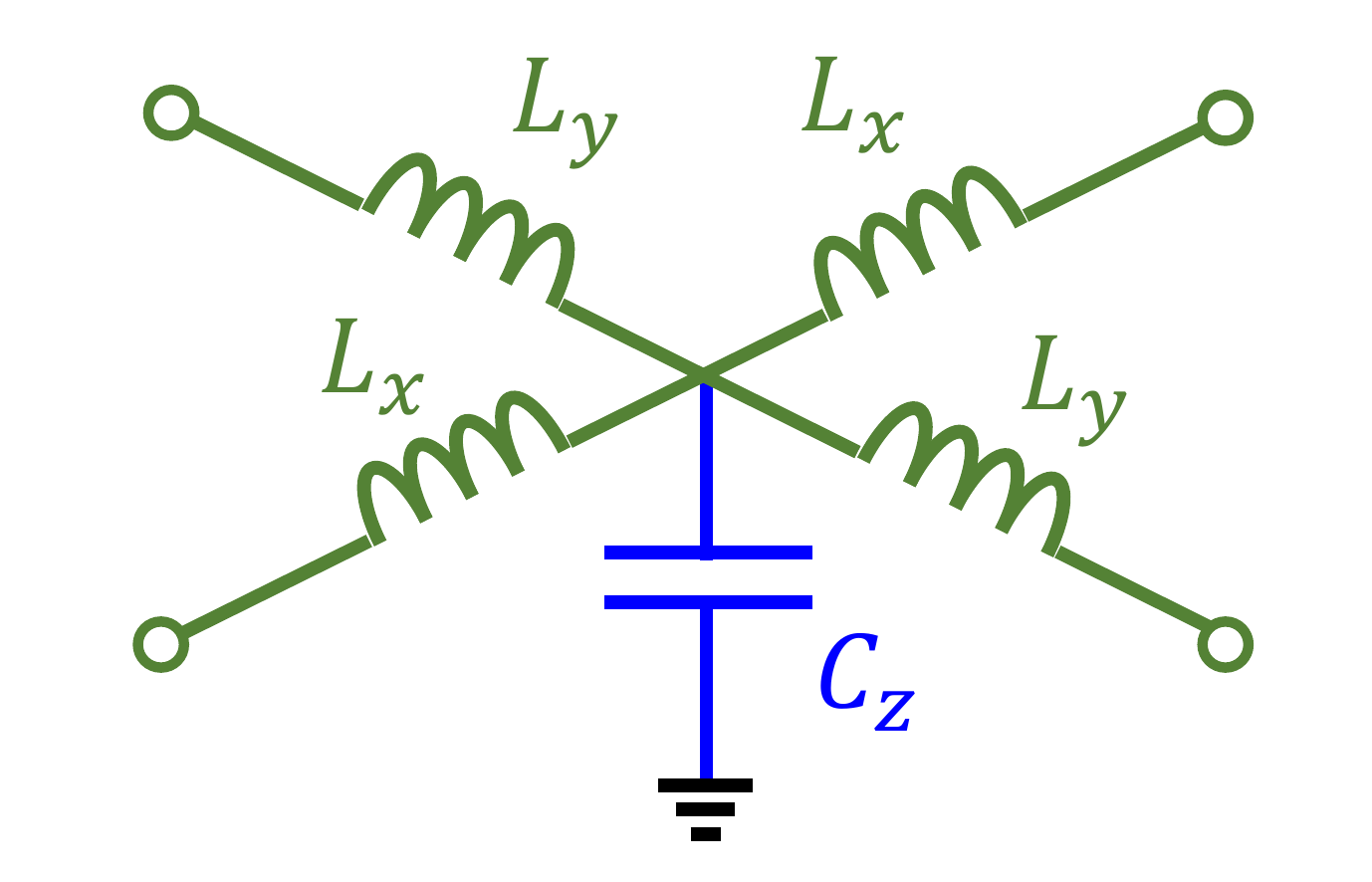}
\caption{{\bf Unit cell circuit model} for an anisotropic transmission line metamaterial (TL-MTM), with effective material parameters: 
$\boldsymbol{\mu}=(\widehat{\mathbf{x}}\widehat{\mathbf{x}}\,L_x +\widehat{\mathbf{y}}\widehat{\mathbf{y}}\,L_y)/\Delta x$ and $\varepsilon_z = C_z/\Delta x$.}
\label{fig:Unit_cell}
\end{figure}

An anisotropic temporal boundary is schematically depicted in Fig.\,\ref{fig:Fig1}, where we focus on the case of two-dimensional (2D) media with isotropic permittivity and anisotropic permeability, considering the behavior of the modes with out-of-plane electric field polarization. Thus, at the temporal boundary $t=t_0$ the system suddenly changes from constitutive parameters $\varepsilon_{z1}$ and $\boldsymbol{\mu}_1$ for $t<t_0$ to $\varepsilon_{z2}$ and $\boldsymbol{\mu}_2$ for $t>t_0$. This material system has been selected because it can be effectively implemented, for example, with the simple 2D TL-MTM with the circuit unit-cell depicted in Fig.\,\ref{fig:Unit_cell} \cite{Caloz2005,Eleftheriades2005negative}. However, we note that, by duality, all effects discussed in this work can be observed in systems with an anisotropic permittivity and out-of-plane magnetic field polarization, which might be more convenient at optical frequencies. Extensions to configurations where both permittivity and permeability are anisotropic are left for future research.

At the temporal boundary, the system suddenly changes its Hamiltonian from $\widehat{H}_1$ to $\widehat{H}_2$, with $\widehat{H}_n=\sum_{\bf k} \hslash\omega_{{\bf k}n}\,\widehat{a}_{{\bf k}n}^{\dagger}\widehat{a}_{{\bf k}n}+1/2$, $n=1,2$, where $\omega_{{\bf k}n}$ is the frequency and $\widehat{a}_{{\bf k}n}$ is the destruction operator for an optical mode with wavevector 
${\bf k}=k({\bf u}_x\cos\phi + {\bf u}_y\sin\phi)$. A crucial aspect of anisotropic media is that modes with the same wavenumber $k$ have a different frequency $\omega_{{\bf k}n}$ as a function of the direction of propagation $\phi$. For example,  
$\omega_{{\bf k}n}=kc\,\sqrt{(\mu_{xn}\mathrm{cos}^{2}\phi+\mu_{y}\mathrm{sin}^{2}\phi)/(\varepsilon_{zn}\mu_{xn}\mu_{yn}})$ for uniaxial media (See Supplementary Information).

Due to the instantaneous nature of the temporal boundary, the response of the system to the change of the Hamiltonian can be modeled within the sudden approximation \cite{Sakurai2014modern}, where the state of the system does not have time enough to follow the changes in the Hamiltonian. Such approximation is justified by taking the asymptotic limit of the time evolution operator:
$\lim_{t\rightarrow t_0}\widehat{U}(t,t_0)=\widehat{I}$ \cite{Sakurai2014modern}, or by integrating Schrödinger's equation around the temporal boundary: 
$\partial_t\left|\psi\right\rangle=(i\hslash)^{-1}\widehat{H}\left|\psi\right\rangle$.

However, because the basis in which the quantum state is written before and after the temporal boundary are different, there are nontrivial changes in the photon statistics. The additional boundary conditions required to compute such photon statistics can be casted in the form of operator transformation rules. Such transformation rules can be found by noting that the quantized field operators obey Maxwell equations: 
$\partial_{t}\mathbf{\widehat{D}}=\nabla\times\mathbf{\widehat{H}}$ 
and 
$\partial_{t}\mathbf{\widehat{B}}=-\nabla\times\mathbf{\widehat{E}}$. Therefore, similar to the classical case, the electric displacement and magnetic flux operators must be continuous across the temporal boundary, i.e.,  
$\mathbf{\widehat{D}}(t=t_0^{-})=\mathbf{\widehat{D}}(t=t_0^{+})$
and
$\mathbf{\widehat{B}}(t=t_0^{-})=\mathbf{\widehat{B}}(t=t_0^{+})$.

Following the quantization of the electromagnetic field in anisotropic media (see Supplementary Information), the magnetic field operator in the interaction picture is given by
$\mathbf{\widehat{H}}\left(\mathbf{r},t\right)=\mathbf{\widehat{H}}^{\left(+\right)}\left(\mathbf{r},t\right)+h.c.$
with positive frequency component 
$\mathbf{\widehat{H}}^{\left(+\right)}\left(\mathbf{r},t\right)=\sum_{{\bf k}}\,\mathbf{\widehat{H}}_{{\bf k}}^{\left(+\right)}\left(\mathbf{r},t\right)$
and individual mode operator
\begin{equation}
\mathbf{\widehat{H}}_{{\bf k}}^{\left(+\right)}\left(\mathbf{r},t\right)=\sqrt{\frac{\hslash\omega_{{\bf k}}}{2\mu_{0}C_{{\bf k}}V}\,}\,\,
\mathbf{h}_{{\bf k}}\,\,\widehat{a}_{{\bf k}}e^{i\mathbf{k}\cdot\mathbf{r}}
e^{-i\omega_{\bf k}t}
\label{eq:Hk}
\end{equation}

\noindent where $\mathbf{h}_{{\bf k}}$ is a unit vector describing the polarization of the magnetic field, $V$ is the quantization volume, and we have defined an energy normalization constant $C_{{\bf k}}=\mathbf{h}_{{\bf k}}\cdot\boldsymbol{\mu}\cdot\mathbf{h}_{{\bf k}}$ (see Supplementary Information).

Using (\ref{eq:Hk}) to enforce the boundary conditions on the electric displacement and magnetic flux operators, we find the following input-output transformation rule for the photonic operators before and after the temporal boundary 
\begin{equation}
\widehat{a}_{{\bf k}2}=\mathrm{cosh}\left(s_{\bf k}\right)\,\widehat{a}_{{\bf k}1}-\mathrm{sinh}\left(s_{\bf k}\right)\,\widehat{a}_{-{\bf k}1}^{\dagger}
\label{eq:input_output}
\end{equation}

\noindent In other words, the input-output relations are a two-mode squeezing transformation with squeezing parameter
\begin{equation}
s_{\bf k}=\mathrm{ln}\sqrt{\frac{\omega_{{\bf k}2}\varepsilon_{z2}}{\omega_{{\bf k}1}\varepsilon_{z1}}}
\label{eq:s_param}
\end{equation}

Previous works have identified that isotropic temporal boundaries result in a squeezing transformation \cite{Mendoncca2000quantum}. Here, we demonstrate that the same conclusion can be extended to anisotropic temporal boundaries. In fact, despite the complexity at the field level introduced by a permeability tensor, the squeezing parameter acquires the simple form given by Eq.\,(\ref{eq:s_param}). However, a crucial difference is that, by contrast with isotropic media, the squeezing parameter $s_{\bf k}$ for modes with the same wavenumber changes as a function of the direction of propagation $\phi$. In turn, this property enables the control over the angular properties of photon production.

Simpler expressions for the squeezing parameter are obtained for particular cases. For an isotropic medium, the squeezing parameter reduces to 
$s_{\bf k}=\mathrm{ln}\sqrt{Z_2/Z_1}\,\,\,\,\forall {\bf k}$. Therefore, the squeezing transformation is the same for all directions, and it is solely determined by the contrast between the medium impedances $Z_n=\sqrt{\mu_n/\varepsilon_n}\,\,\,n=1,2$. On the other hand, if the permittivity is the same in both media, $\varepsilon_{z1}=\varepsilon_{z2}$, the squeezing parameter reduces to $s_{\bf k}=\mathrm{ln}\sqrt{\omega_{{\bf k}2}/\omega_{{\bf k}1}}$, revealing that only the frequency shift of the modes results in a squeezing effect. 

\subsection{Vacuum amplification and photon statistics}

Let us assume that the system is initially in the vacuum state $\left|0\right\rangle$, with a zero average on the number of photons 
$\left\langle n_{{\bf k}1}\right\rangle=\left\langle a_{{\bf k}1}^{\dagger}a_{{\bf k}1}\right\rangle=0$, and minimal variance in the quadrature operators: $\Delta X_{{\bf k}1}^2=\Delta Y_{{\bf k}1}^2=1/4$, with 
$\Delta A_{{\bf k}}^2=\left\langle\widehat{A}_{\bf k}^2\right\rangle - \left\langle\widehat{A}_{\bf k}\right\rangle^2$, 
$\widehat{X}_{\bf k} = (\widehat{a}_{\bf k}^{\dagger} + \widehat{a}_{\bf k})/2$ and
$\widehat{Y}_{\bf k} = i(\widehat{a}_{\bf k}^{\dagger} - \widehat{a}_{\bf k})/2$.
After the temporal boundary, the average number of photons is $\left\langle n_{{\bf k}2}\right\rangle={\rm sinh}^2(s_{\bf k})$, and the variances of the quadrature operators are 
$\Delta X_{{\bf k}2}^2=\Delta Y_{{\bf k}2}^2=(1+{\rm sinh}^2(s_{\bf k}))/2$. As expected, the temporal boundary induces a vacuum amplification effect, generating photons from the electromagnetic vacuum state. In addition, the larger the squeezing parameter $s_{\bf k}$ the larger the photon production is, and the larger the variances of quadrature operators are.

\begin{figure}[!t]
\includegraphics[width=3.35in]{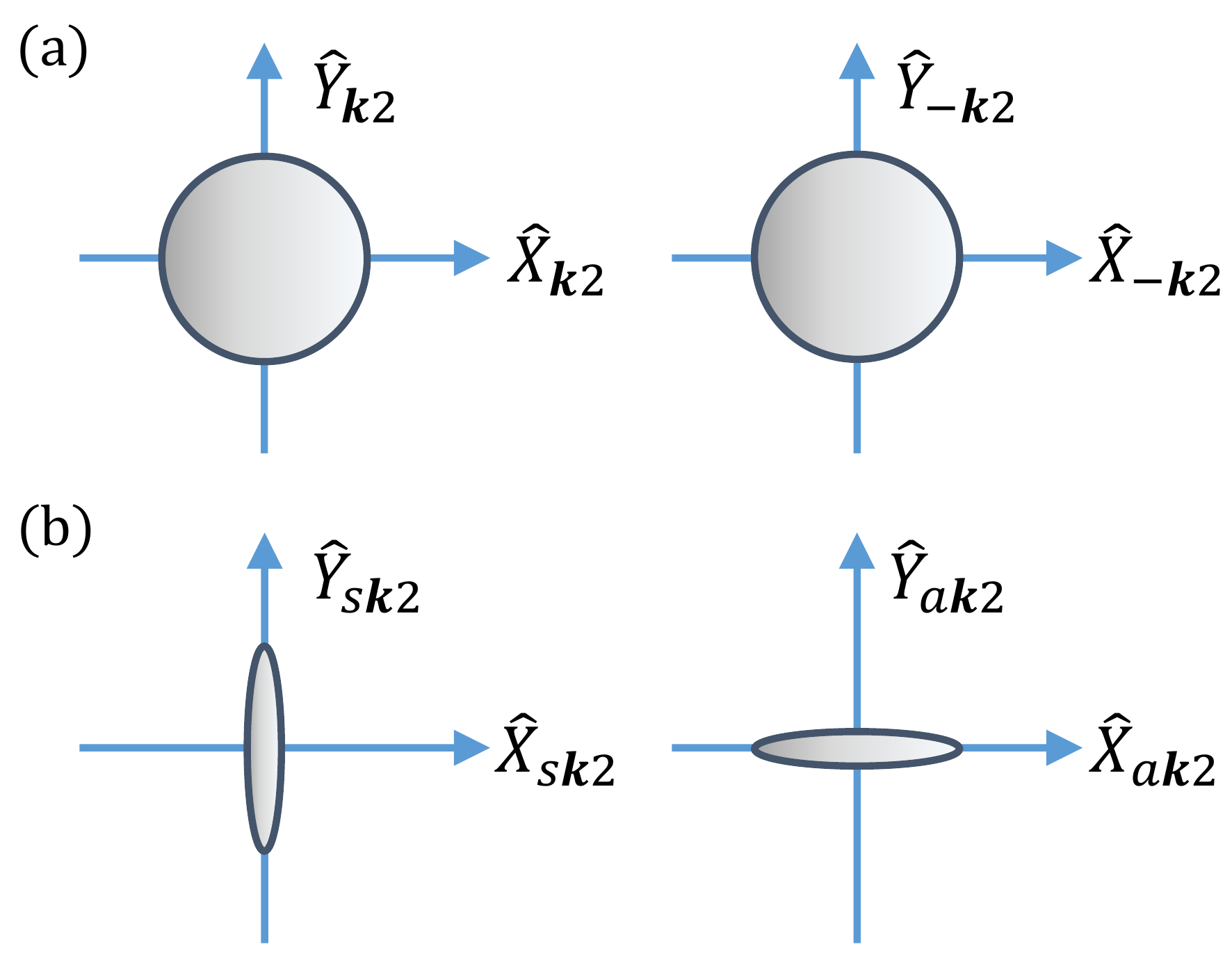}
\caption{{\bf Photon statistics after the temporal boundary.} (a) Schematic representation of the quadrature operator variances for the forward (left) and backward (right) modes with wavenumber ${\bf k}$. (b) Same as in (a), but in the symmetric (left) and asymmetric (b) basis, showing squeezing along different axis.}
\label{fig:Photon_statistics}
\end{figure}

The squeezing nature of the temporal boundary is more apparent when the output modes are analyzed in a symmetric/anti-symmetric basis:
$\widehat{a}_{s{\bf k}2} = (\widehat{a}_{{\bf k}2} + \widehat{a}_{-{\bf k}2})/\sqrt{2}$
and
$\widehat{a}_{a{\bf k}2} = (\widehat{a}_{{\bf k}2} - \widehat{a}_{-{\bf k}2})/\sqrt{2}$.
In this basis, the quadrature variances are exponentially expanded/compressed with the squeezing operator:
$\Delta X_{s{\bf k}2}^2=\Delta Y_{a{\bf k}2}^2=e^{-2s_{\bf k}}/4$
and
$\Delta Y_{s{\bf k}2}^2=\Delta X_{a{\bf k}2}^2=e^{2s_{\bf k}}/4$, as schematically depicted in Fig.\,\ref{fig:Photon_statistics}. In conclusion, anisotropic temporal boundaries result in photon generation from vacuum, with nontrivial correlations between the photons propagating along opposite directions.

\subsection{Angular-dependent photon production}

\begin{figure}[!t]
\includegraphics[width=3.35in]{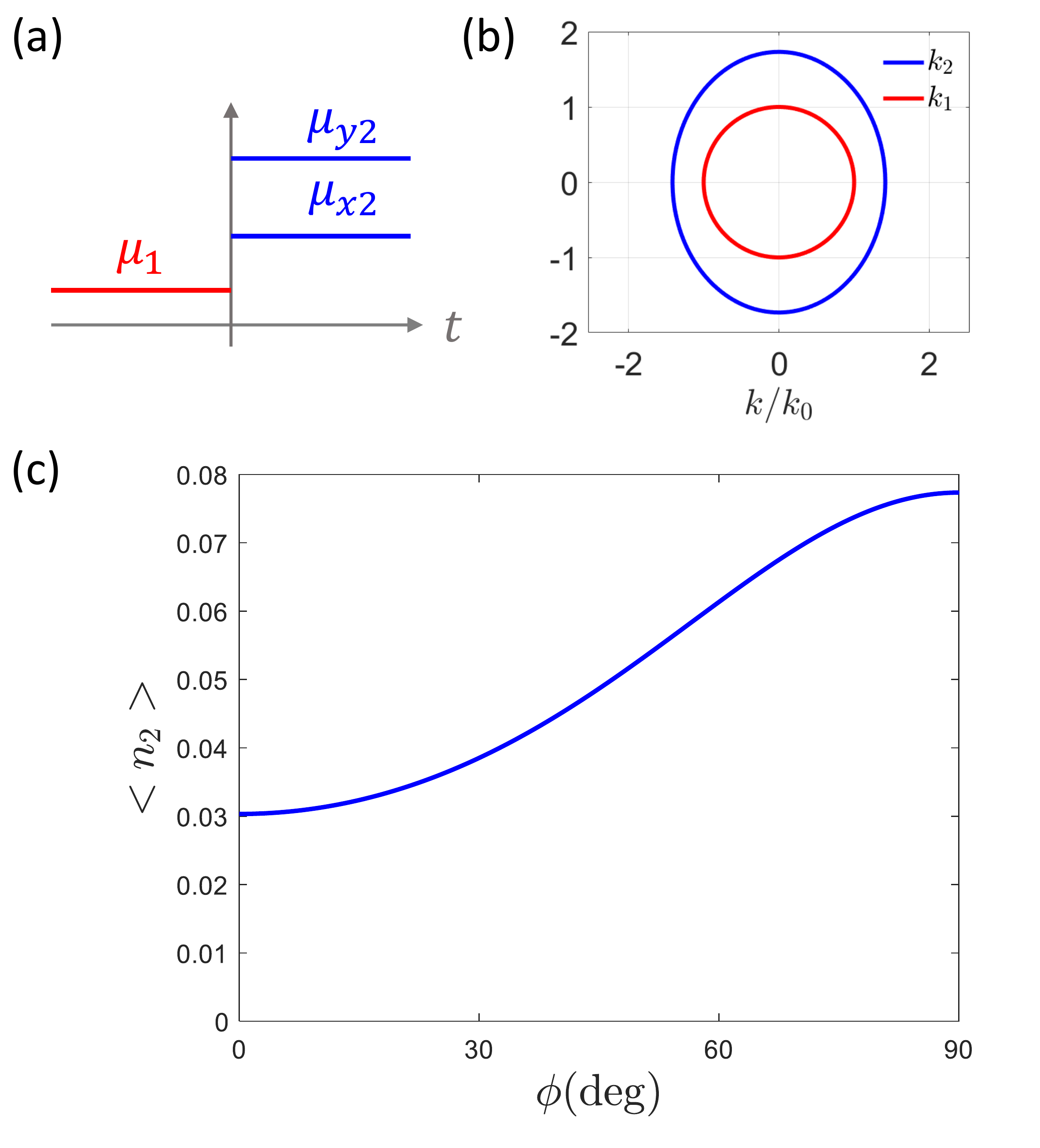}
\caption{{\bf Angular distribution of the photon production}. (a) Sketch of an anisotropic temporal boundary, where an isotropic medium, 
$\boldsymbol{\mu}_1=\mu_1\,(\widehat{\bf{x}}\widehat{\bf{x}}+\widehat{\bf{y}}\widehat{\bf{y}})$, is suddenly transformed into an 
anisotropic medium, $\boldsymbol{\mu}_2=\mu_{2x}\,\widehat{\bf{x}}\widehat{\bf{x}}+\mu_{2y}\,\widehat{\bf{y}}\widehat{\bf{y}}$. The pemittivity is assumed to be unity in both cases, $\varepsilon_{1z}=\varepsilon_{2z}=1$. (b) Isofrequency contours before and after the temporal boundary, for $\mu_1=1$, $\mu_{2x}=2$ and $\mu_{2y}=3$. (c) Photon production for modes with a common wavenumber $k$ as a function of the angle of propagation $\phi$.}
\label{fig:Single_1}
\end{figure}

The main signature of quantum anisotropic temporal boundaries is the angular dependence in photon production. To illustrate this point, we apply our theory to the particular case schematically depicted in Fig.\,\ref{fig:Single_1}(a), where a medium is isotropic before the temporal boundary,  
$\boldsymbol{\mu}_1=\mu_1\,(\widehat{\bf{x}}\widehat{\bf{x}}+\widehat{\bf{y}}\widehat{\bf{y}})$, while it changes to an anisotropic medium, with a diagonal permeability tensor,
$\boldsymbol{\mu}_2=\mu_{2x}\,\widehat{\bf{x}}\widehat{\bf{x}}+\mu_{2y}\,\widehat{\bf{y}}\widehat{\bf{y}}$, after the temporal boundary. A unit-permittivity medium is assumed before and after the temporal boundary, $\varepsilon_{z1}=\varepsilon_{z2}=1$. Consequently, the isofrequency contours, depicted in Fig.\,\ref{fig:Single_1}(b) for the particular case of $\mu_1=1$, $\mu_{2x}=2$ and $\mu_{2y}=3$, shift from circular to elliptical across the temporal boundary. The angular-dependent photon production is depicted in Fig.\,\ref{fig:Single_1}(c), confirming that the number of generated photons presents a clear angular dependence as a result of anisotropy of the temporal boundary. In particular, the photon production smoothly varies between the values associated with the permeability contrast on each of the ellipse semiaxis. 

\subsection{Inhibiting vacuum amplification}

\begin{figure}[!t]
\includegraphics[width=3.35in]{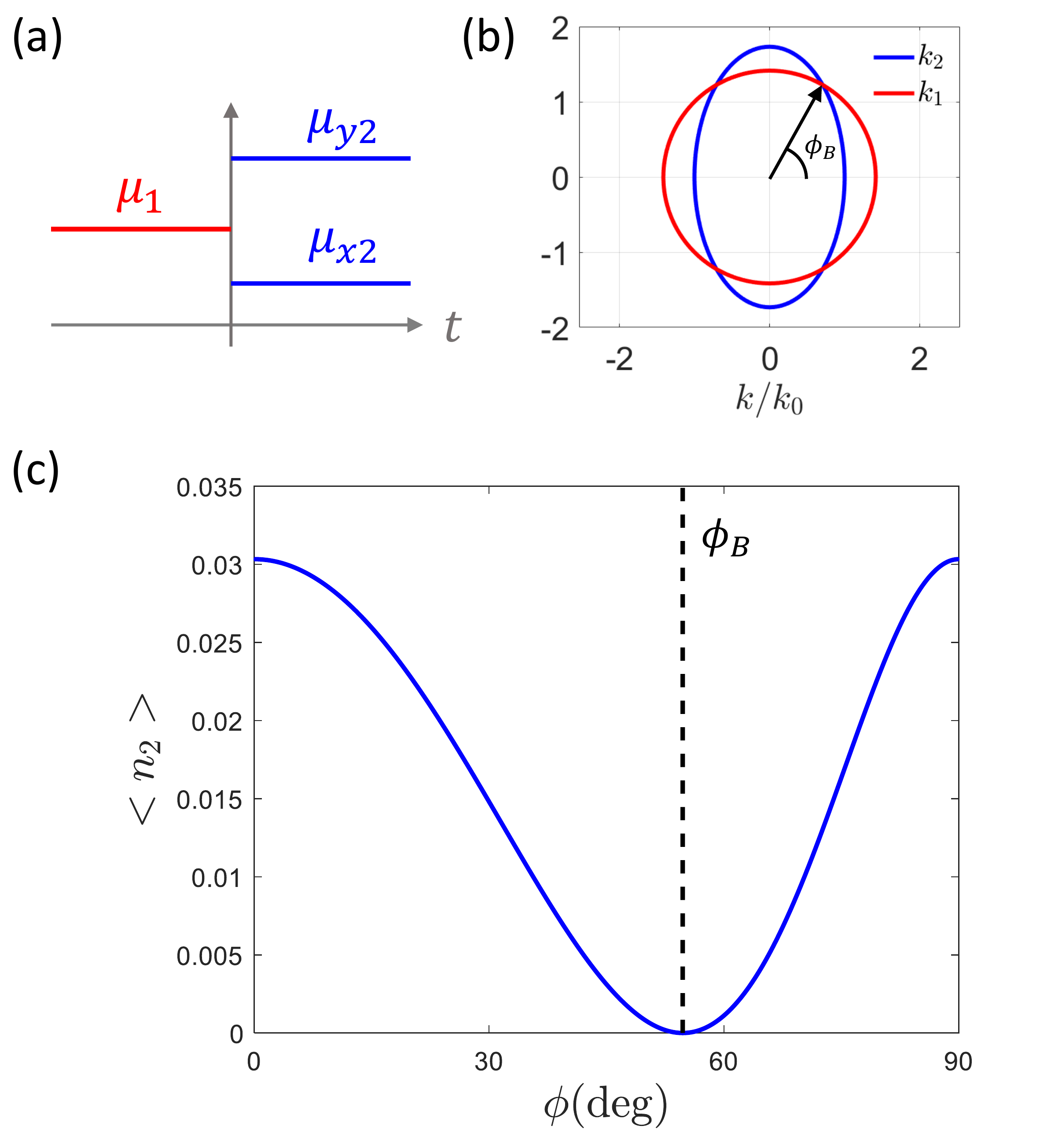}
\caption{{\bf Anisotropic temporal boundaries with intersecting isofrequencies}. (a) Sketch of an anisotropic temporal boundary, where an isotropic medium, 
$\boldsymbol{\mu}_1=\mu_1\,(\widehat{\bf{x}}\widehat{\bf{x}}+\widehat{\bf{y}}\widehat{\bf{y}})$, is suddenly transformed into an 
anisotropic medium, $\boldsymbol{\mu}_2=\mu_{2x}\,\widehat{\bf{x}}\widehat{\bf{x}}+\mu_{2y}\,\widehat{\bf{y}}\widehat{\bf{y}}$. The pemittivity is assumed to be unity in both cases, $\varepsilon_{1z}=\varepsilon_{2z}=1$. (b) Isofrequency contours before and after the temporal boundary, for $\mu_1=2$, $\mu_{2x}=1$ and $\mu_{2y}=4$. (c) Photon production for modes with a common wavenumber $k$ as a function of the angle of propagation $\phi$. The dashed black line indicates the angle $\sin\phi_B=\sqrt{\mu_{y2}(\mu_{x2}-\mu_{1})/\mu_{1}(\mu_{x2}-\mu_{y2})}$ where the isofrequency contours cut, corresponding to a zero of photon production.}
\label{fig:Single_2}
\end{figure}

A more interesting angular distribution of photon production is found when the isofrequencies of the media on both sides of the temporal boundary intersect. For example, Fig.\,\ref{fig:Single_2}(a) shows an anisotropic temporal boundary with material parameters $\mu_1=2$, $\mu_{2x}=1$ and $\mu_{2y}=4$, such that the isotropic permeability value before the temporal boundary lies between the two diagonal values of the anisotropic case, i.e., $\mu_{2x}<\mu_{1}<\mu_{2y}$. In this manner, the circular isofrecuency contour before the temporal boundary crosses the elliptical isofrequency contour after the temporal boundary (see Fig.\,\ref{fig:Single_2}(b)). 

The average number of generated photons for this configuration is depicted in Fig.\,\ref{fig:Single_2}(c). The angular distribution is characterized by a zero at the angle where the isofrequencies cut, given by the solution to
$\sin\phi_B=\sqrt{\mu_{y2}(\mu_{x2}-\mu_{1})/\mu_{1}(\mu_{x2}-\mu_{y2})}$. At this angle, both media support a mode with the same wavenumber and at the same frequency, so that no frequency shift is needed to support the boundary conditions across the temporal boundary. Following Eq.\,(\ref{eq:s_param}), the squeezing parameter is zero at such angle. This effect is the quantum counterpart to the classical temporal Brewster angle derived in \cite{pacheco2021Brewster}, where it was found that the field reflected at the temporal boundary is cancelled for specific angles. Here, it is demonstrated that tailoring the anisotropy of the temporal boundary it is possible to inhibit vacuum amplification effects at specific angles. 

\section{Multilayered anisotropic temporal boundaries}

Additional design flexibility can be obtained by combining multiple temporal boundaries to conform a temporal sequence. In this section, we demonstrate that multilayered anisotropic temporal boundaries enable vacuum amplification effects while: (i) concentrating all generated photons into a single direction, (ii) generating angular and frequency combs of photons, (iii) enforcing asymmetric amplification lines with fast angular variation. 

\subsection{Theoretical framework}

First, we generalize the theory introduced before for the case of multiple temporal boundaries. To this end, it is convenient to rewrite the input-output relations in the form of a transfer matrix:
\begin{equation}
\left[\begin{array}{c}
\widehat{a}_{{\bf k}2}\left(t\right)\\
\widehat{a}_{-{\bf k}2}^{\dagger}\left(t\right)
\end{array}\right]=\boldsymbol{\mathcal{S}}\left(s_{\bf k}\right)\left[\begin{array}{c}
\widehat{a}_{{\bf k}1}\left(t\right)\\
\widehat{a}_{-{\bf k}1}^{\dagger}\left(t\right)
\end{array}\right]
\label{eq:S_io}
\end{equation}

\noindent with the squeezing matrix
\begin{equation}
\boldsymbol{\mathcal{S}}\left(s_{\bf k}\right)=\left[\begin{array}{cc}
\mathrm{cosh}\left(s_{\bf k}\right) & -\mathrm{sinh}\left(s_{\bf k}\right)\\
\mathrm{-sinh}\left(s_{\bf k}\right) & \mathrm{cosh}\left(s_{\bf k}\right)
\end{array}\right]
\label{eq:S_matrix}
\end{equation}

Similarly, the time evolution along a temporal slab can be characterized in transfer matrix form as
\begin{equation}
\left[\begin{array}{c}
\widehat{a}_{{\bf k}}\left(t+\tau\right)\\
\widehat{a}_{-{\bf k}}^{\dagger}\left(t+\tau\right)
\end{array}\right]=\mathbf{U}\left(\omega\tau\right)\left[\begin{array}{c}
\widehat{a}_{{\bf k}}\left(t\right)\\
\widehat{a}_{-{\bf k}}^{\dagger}\left(t\right)
\end{array}\right]
\label{eq:U_io}
\end{equation}

\noindent with the time-evolution matrix
\begin{equation}
\mathbf{U}\left(\varphi\right)=\left[\begin{array}{cc}
e^{-i\varphi} & 0\\
0 & e^{i\varphi}
\end{array}\right]
\label{eq:U_matrix}
\end{equation}

Consequently, the input-output relations for a sequence of $N-1$ anisotropic temporal boundaries (thus with $N$ temporal layers) can be compactly written as 
\[
\left[\begin{array}{c}
\widehat{a}_{kN}\\
\widehat{a}_{-kN}^{\dagger}
\end{array}\right]=
\boldsymbol{\mathcal{S}}\left(s_{NN-1}\right)\mathbf{U}\left(\varphi_{N-1}\right)\cdots
\]
\begin{equation}
\cdots\boldsymbol{\mathcal{S}}\left(s_{32}\right)\mathbf{U}\left(\varphi_{2}\right)\boldsymbol{\mathcal{S}}\left(s_{21}\right)\left[\begin{array}{c}
\widehat{a}_{k1}\\
\widehat{a}_{-k1}^{\dagger}
\end{array}\right]
\label{eq:N_boundaries}
\end{equation}

The squeezing and time evolution matrices have nice multiplicative properties,
$\boldsymbol{\mathcal{S}}\left(s_{1}\right)\boldsymbol{\mathcal{S}}\left(s_{2}\right)=\boldsymbol{\mathcal{S}}\left(s_{1}+s_{2}\right)$
and 
$\mathbf{U}\left(\varphi_{1}\right)\mathbf{U}\left(\varphi_{2}\right)=\mathbf{U}\left(\varphi_{1}+\varphi_{2}\right)$
 showing that consecutive squeezing and/or time evolution processes result in the addition of the squeezing parameters and/or phase delays, respectively. However, the same properties do not apply for multilayered slabs, where squeezing and time evolution matrices appear alternating each other. Exceptions occur for specific durations of the temporal slabs. A particularly relevant case for coherent amplification is that of a $T/4$ slab, with $T=2\pi/\omega$ (i.e., one for which the duration of the temporal satisfies $\omega\tau=\pi/2$), for which we can write
\begin{equation}
\boldsymbol{\mathcal{S}}\left(s_{2}\right)\mathbf{U}\left(\frac{\pi}{2}\right)\boldsymbol{\mathcal{S}}\left(s_{1}\right)=\boldsymbol{\mathcal{S}}\left(s_{2}-s_{1}\right)\mathbf{U}\left(\frac{\pi}{2}\right)
\end{equation}

Consequently, the input-output relations for the $T/4$ slab can be factorized as
\[
\left[\begin{array}{c}
\widehat{a}_{{\bf k}N}\\
\widehat{a}_{-{\bf k}N}^{\dagger}
\end{array}\right]=
\]
\[
\boldsymbol{\mathcal{S}}\left(\sum_{n=2}^{N}\left(-1\right)^{N-n}\,s_{{\bf k}nn-1}\right)
\]
\begin{equation}
\,\mathbf{U}\left((N-2)\frac{\pi}{2}\right)\left[\begin{array}{c}
\widehat{a}_{{\bf k}1}\\
\widehat{a}_{-{\bf k}1}^{\dagger}
\end{array}\right]
\label{eq:io_T4}
\end{equation}


\begin{figure*}[!t]
\includegraphics[width=7.0in]{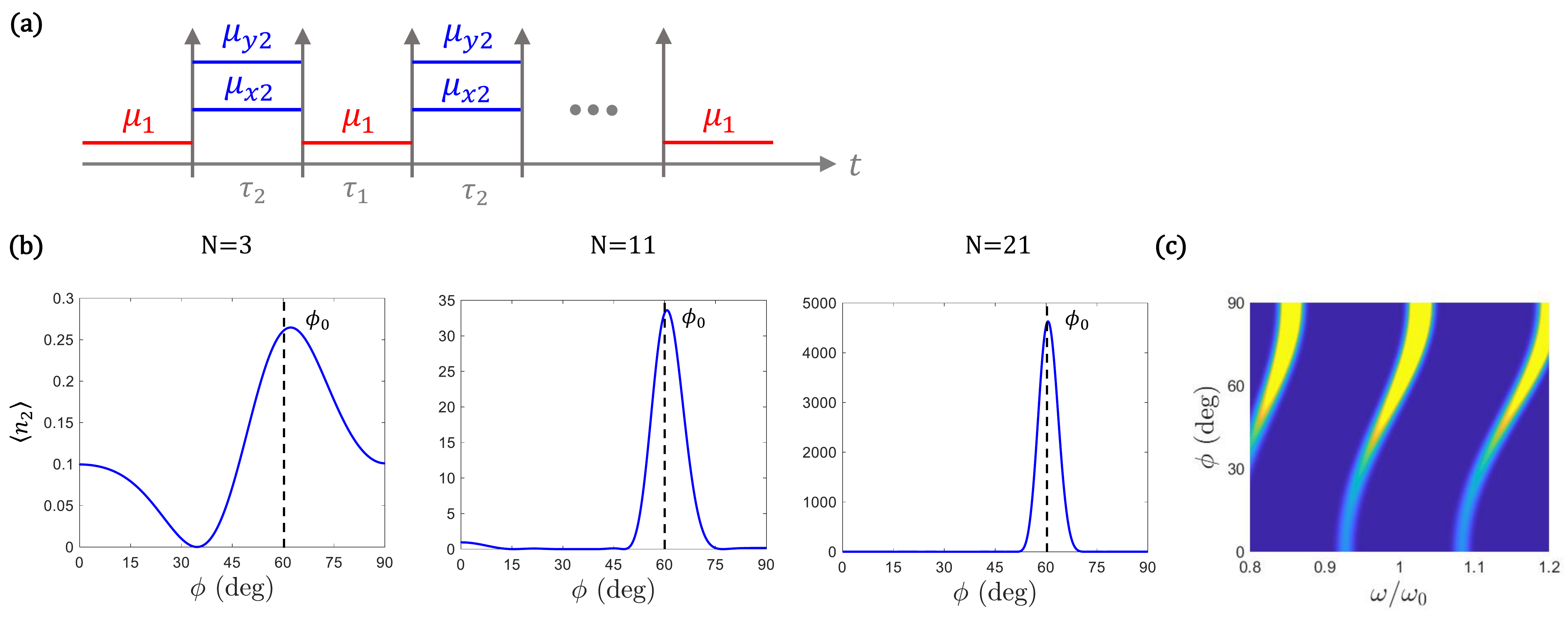}
\caption{{\bf Vacuum amplification concentrated on a single direction.} (a) Sketch of a $N$-layered temporal boundary sequence constructed by the concatenation of unit-permittivity ($\varepsilon_{z1}=\varepsilon_{z2}=1$) isotropic temporal slabs $\boldsymbol{\mu}_1=\mu_1\,(\widehat{\bf{x}}\widehat{\bf{x}}+\widehat{\bf{y}}\widehat{\bf{y}})$ of duration $\tau_1$, and anisotropic temporal slabs 
$\boldsymbol{\mu}_2=\mu_{2x}\,\widehat{\bf{x}}\widehat{\bf{x}}+\mu_{2y}\,\widehat{\bf{y}}\widehat{\bf{y}}$ of duration $\tau_2$. 
(b) Photon production from a mode with wavenumber $k$ as a function of the observation angle $\phi$ for temporal sequences of $N=3$, $N=11$ and $N=21$ layers, with $\mu_1=1$, $\mu_{2x}=2$ and $\mu_{2y}=3$, 
$\tau_1=(\pi/2)/(kc)$ and 
$\tau_2=(5\pi+\pi/2)/\omega_{{\bf k}2}(\phi_0)$, with $\phi_0=60\,{\rm deg}$. The results show that the photon production exponentially grow with the number of temporal layers, while being concentrated along the $\phi_0$ direction. (c) Colormap of the photon production as a function of observation angle $\phi$ and frequency $\omega'/\omega_0$, with $\omega'=k'c$ and $\omega_0=kc$, showing that the peak angle periodically shifts with frequency scanning all directions.}
\label{fig:Amp_one_direction}
\end{figure*}


\subsection{Resonantly-enhanced vacuum amplification along a single direction}

Next we show that multilayered anisotropic slabs can be designed to produce resonantly-enhanced vacuum amplification, while concentrating all generated photons into a single direction. To this end, we consider the two-stage $N$-layered temporal boundary sequence shown in Fig.\,\ref{fig:Amp_one_direction}(a), constructed by the concatenation of two classes of temporal slabs: isotropic temporal slabs $\boldsymbol{\mu}_1=\mu_1\,(\widehat{\bf{x}}\widehat{\bf{x}}+\widehat{\bf{y}}\widehat{\bf{y}})$ of duration $\tau_1$, and  anisotropic temporal slabs 
$\boldsymbol{\mu}_2=\mu_{2x}\,\widehat{\bf{x}}\widehat{\bf{x}}+\mu_{2y}\,\widehat{\bf{y}}\widehat{\bf{y}}$ of duration $\tau_2$ (both with $\varepsilon_{z1}=\varepsilon_{z2}=1$). A two-stage sequence can be convenient from a practical standpoint. In addition, it has the important property that each temporal boundary is the reverse process of their neighbouring boundaries. Consequently, in accordance to Eq.\,(\ref{eq:s_param}) their squeezing parameters only differ by a minus sign 
\begin{equation}
s_{{\bf k}nn-1}=-s_{{\bf k}n+1n}\,\,\,\,\forall n
\label{eq:s_k_property}
\end{equation}

To guide our thoughts, consider the set of modes with wavenumber $k$ but different propagation angle $\phi$. Before the first temporal boundary, the medium is isotropic. Therefore, all modes with the same wavenumber $k$ exist at the same frequency $\omega_{{\bf k}1}=kc/\sqrt{\mu_1}$ for all directions of propagation $\phi$. After the temporal boundary, the medium is anisotropic, and each mode is projected into a different frequency as a function of the angle of propagation, 
$\omega_{{\bf k}2}(\phi)=kc\sqrt{\mu_{2x}\mu_{2y}/(\mu_{2x}\cos^2\phi + \mu_{2y}\sin^2\phi)}$
, i.e., the anisotropic temporal boundary behaves as an ``inverse prism" \citep{akbarzadeh2018inverse}. In this manner, the phase advance $\omega_{{\bf k}2}\tau_2$ experienced by each mode changes as a function of its direction of propagation. After the next temporal boundary, the medium becomes isotropic again, and all modes converge to the same frequency $\omega_{{\bf k}1}$. As this process is repeated multiple times, the temporal sequence experienced by each mode is different, due to their different phase advances through medium 2. Therefore, when the temporal sequence finishes, all modes are back to the original $\omega_{{\bf k}1}$ frequency. However, their photon statistics change in a nontrivial fashion as a function of the angle of propagation, following Eq.\,(\ref{eq:N_boundaries}).

Let us assume that there is a direction $\phi_0$, with wavevector ${\bf k}_0$, for which the time-duration of the temporal slabs is $T/4$ resonant. That is to say, the duration of the time delays is set in such a way that $\omega_{{\bf k}_01}\tau_1=n_1\pi + \pi/2$ and $\omega_{{\bf k}_02}\tau_2=n_2\pi + \pi/2$. In this case, Eq.\,(\ref{eq:N_boundaries}) simplifies to (\ref{eq:io_T4}), and using (\ref{eq:s_k_property}) it further reduces to
\begin{equation}
\left[\begin{array}{c}
\widehat{a}_{{\bf k}_0N}\\
\widehat{a}_{-{\bf k}_0N}^{\dagger}
\end{array}\right]=\boldsymbol{\mathcal{S}}\left((N-1)s_{{\bf k}0}\right)\,\mathbf{U}\left((N-2)\frac{\pi}{2}\right)\left[\begin{array}{c}
\widehat{a}_{{\bf k}_01}\\
\widehat{a}_{-{\bf k}_01}^{\dagger}
\end{array}\right]
\label{eq:io_resonant}
\end{equation}

In other words, for the mode with wavevector ${\bf k}_0$, associated with the direction $\phi_0$, the temporal sequence acts as a squeezing transfomation with enhanced squeezing parameter $(N-1)s_{{\bf k}_0}$. Thus, the photon production from the quantum vacuum in the $\phi_0$ direction is coherently amplified. However, the same resonant condition does not hold for the rest of $\phi$ directions. 

In order to illustrate this effect, Fig.\,\ref{fig:Amp_one_direction} represents the photon production from a mode with wavenumber $k$ as a function of the observation angle $\phi$ for temporal sequences of $N=3$, $N=11$ and $N=21$ layers, with $\mu_1=1$, $\mu_{2x}=2$ and $\mu_{2y}=3$, 
$\tau_1=(\pi/2)/(kc)$ and 
$\tau_2=(5\pi+\pi/2)/\omega_{{\bf k}_02}$, with $\phi_0=60\,{\rm deg}$. It can be concluded from the figure that the photon production exponentially grows with the number of temporal layers, while being concentrated along the $\phi_0$ direction. Therefore, it is shown that by using anisotropic temporal boundaries it is possible to resonantly generate photons from the quantum vacuum, while concentrating all of them along a specific direction.


\begin{figure*}[!t]
\includegraphics[width=7.0in]{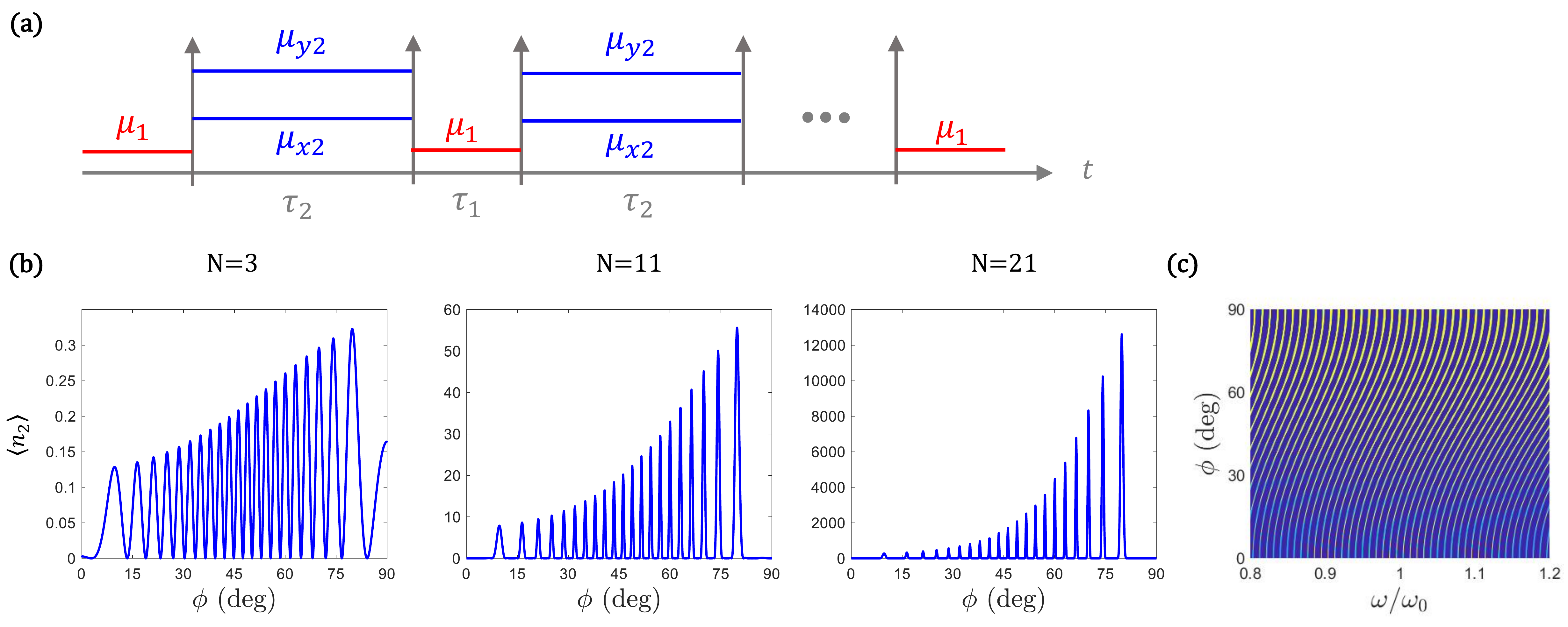}
\caption{{\bf Generation of angular and frequency combs.} Same as in Fig.\,\ref{fig:Amp_one_direction}, except the duration of the second temporal slab has been extended to $\tau_2=(100\pi+\pi/2)/\omega_{{\bf k}_02}$, with $\phi_0=60\,{\rm deg}$. Plots in (b) show that photon yields an angular comb with multiple peaks of emission. Plots in (c) show that the angular comb periodically shiftes with frequency, forming a frequency comb along a single direction.}
\label{fig:Amp_comb}
\end{figure*}


\subsection{Generation of angular and frequency combs}

Controlling the parameters of the temporal sequence enables shaping the photon production beyond concentrating it into a single direction. For example, if the time interval $\tau_2$ is large enough, more than one angle will satisfy the condition for $T/4$ resonant amplification. In such a case, the photon production will be angularly characterized by a comb structure, arising from multiple $\phi$ directions resonating simultaneously. 

We illustrate this effect by considering again the temporal sequence depicted in Fig.\,\ref{fig:Amp_one_direction}(a), but extending the duration of the anisotropic time slabs to $\tau_2=(100\pi+\pi/2)/\omega_{{\bf k}_02}$, as schematically depicted in Fig.\,\ref{fig:Amp_comb}(a). As expected, the photon production for a mode with wavenumber $k$, shown in Fig.\,\ref{fig:Amp_comb}(b), exhibits an angular comb with multiple amplification peaks. The envelope of the comb is a result of the response of each individual temporal boundary (see Fig.\,\ref{fig:Single_1}), and it is associated with the permeability contrast along different directions. 

Photon production as a function of frequency and angle of propagation is reported in Fig.\,\ref{fig:Amp_comb}(c) where it is shown that each amplification peak continuously shifts its angle with frequency. As a result, photon production exhibits an angular comb for a fixed frequency, but also a frequency comb for a fixed angle of observation. 


\begin{figure}[!t]
\includegraphics[width=3.25in]{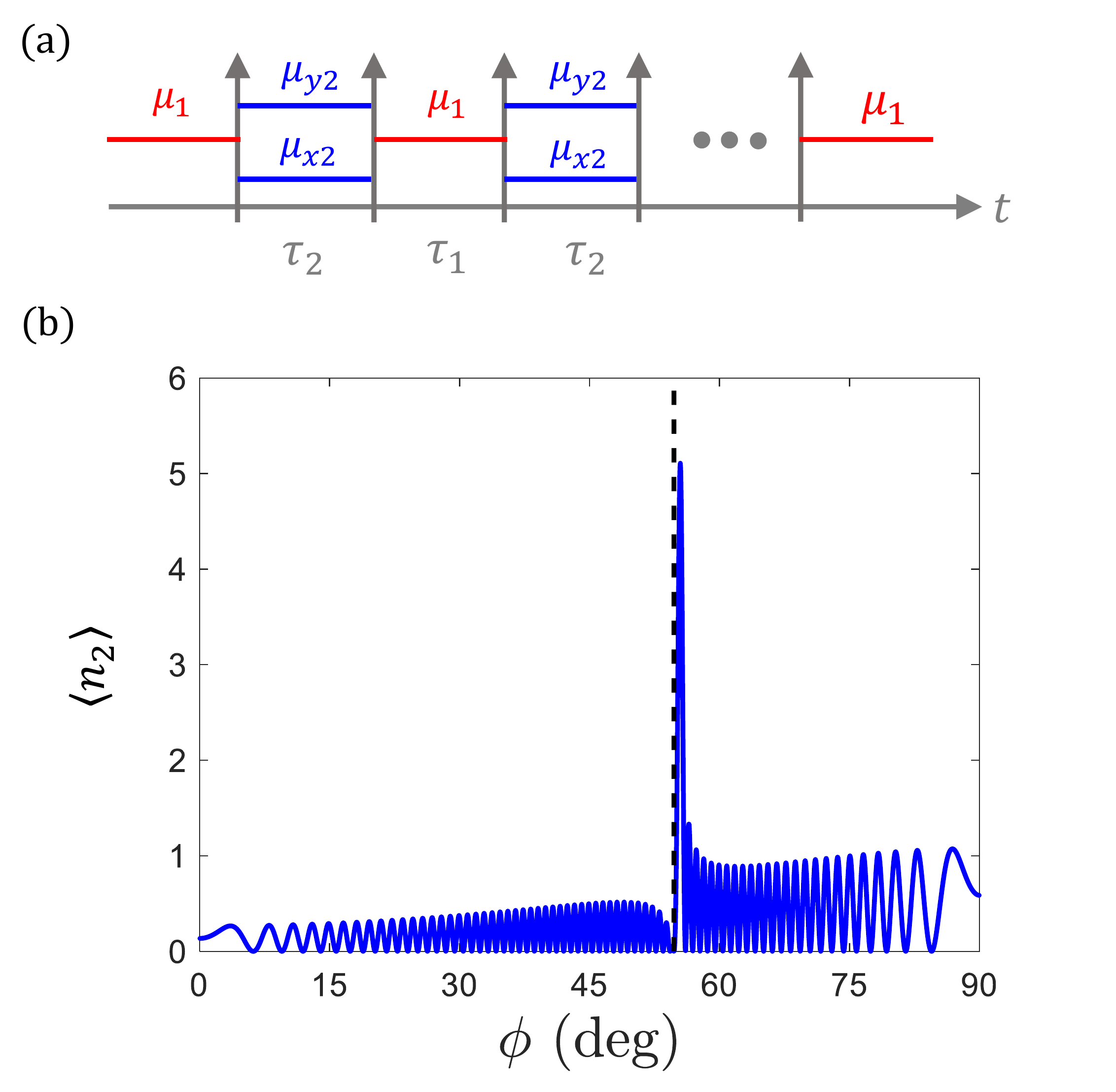}
\caption{{\bf Asymmetric and fast-varying angular amplification lines near a temporal Brewster angle.} (a) Sketch of a temporal sequence constructed by 501 temporal slabs with material parameters: $\varepsilon_{z1}=\varepsilon_{z2}=1$, $\boldsymbol{\mu}_1=2\,(\widehat{\bf{x}}\widehat{\bf{x}}+\widehat{\bf{y}}\widehat{\bf{y}})$  and $\boldsymbol{\mu}_2=1\,\widehat{\bf{x}}\widehat{\bf{x}}+4\,\widehat{\bf{y}}\widehat{\bf{y}}$, and duration $\tau_1=(\pi/2)/(kc)$ and $\tau_2=(\pi/2)/\omega_{{\bf k}_02}$. Inhibition of takes place near $\phi_{B} = 54.74\,{\rm deg}$, while resonant amplification takes place at $\phi_0=55\,{\rm deg}$. 
(b) Photon production as a function of the observation angle $\phi$. The results show an asymmetric angular line arising from the interaction between the Brewster angle and the coherent amplification.}
\label{fig:Amp_brewster}
\end{figure}


\subsection{Asymmetric and fast-varying angular lines}

Different resonant amplification effects can be observed in temporal sequences containing anisotropic boundaries whose isofrequencies intersect. In such cases, the interaction between resonant amplification and the temporal Brewster angle leads to asymmetric angular lines, akin to a Fano resonances. To illustrate this point, we select material parameters $\mu_1=1$, $\mu_{2x}=2$ and $\mu_{2y}=3$ such that vacuum amplification effects are inhibited at the Brewster angle: $\phi_{B} = 54.74\,{\rm deg}$. The duration of the temporal slabs  
$\tau_1=(\pi/2)/(kc)$ and 
$\tau_2=(\pi/2)/\omega_{{\bf k}_02}$ are selected to enforce amplification at $\phi_0=55\,{\rm deg}$. The resulting photon production, shown in Fig.\,(\ref{fig:Amp_brewster}) is characterized by a fast angular variation, in the form of an asymmetric angular line arising from the interaction between the Brewster angle and the resonant amplification (see Fig.\,\ref{fig:Amp_brewster}(b)).

\section{Conclusions}

Our results theoretically demonstrate that anisotropic temporal boundaries provide control over the angular distribution of the photons generated via vacuum amplification effects. We show how the design of the anisotropy of the temporal boundaries and their time sequencing empower the inhibition of the photon production in specific directions, the resonant photon production concentrated along a single direction, the emission of angular and frequency photon combs, and asymmetric photon generation with rapid angular variations. We expect that research over more advanced temporal sequences would lead to even a finer control on the angular distribution and/or the realization of arbitrary angular distributions. In general, the design flexibility offered by temporal metamaterials enriches the physics of quantum optical phenomena. Upon this basis, one can envision further advances of this field where each of the different degrees of freedom of a material connects with a degree of freedom of the generated photons, with their interaction enforcing nontrivial quantum correlations.

\begin{acknowledgments}
I.L. acknowledges support from Ram\'on y Cajal fellowship RYC2018-024123-I and project RTI2018-093714-301J-I00 sponsored by MCIU/AEI/FEDER/UE, and ERC Starting Grant 948504. 
\end{acknowledgments}

\bibliography{library}

\begin{thebibliography}{37}%
\makeatletter
\providecommand \@ifxundefined [1]{%
 \@ifx{#1\undefined}
}%
\providecommand \@ifnum [1]{%
 \ifnum #1\expandafter \@firstoftwo
 \else \expandafter \@secondoftwo
 \fi
}%
\providecommand \@ifx [1]{%
 \ifx #1\expandafter \@firstoftwo
 \else \expandafter \@secondoftwo
 \fi
}%
\providecommand \natexlab [1]{#1}%
\providecommand \enquote  [1]{``#1''}%
\providecommand \bibnamefont  [1]{#1}%
\providecommand \bibfnamefont [1]{#1}%
\providecommand \citenamefont [1]{#1}%
\providecommand \href@noop [0]{\@secondoftwo}%
\providecommand \href [0]{\begingroup \@sanitize@url \@href}%
\providecommand \@href[1]{\@@startlink{#1}\@@href}%
\providecommand \@@href[1]{\endgroup#1\@@endlink}%
\providecommand \@sanitize@url [0]{\catcode `\\12\catcode `\$12\catcode
  `\&12\catcode `\#12\catcode `\^12\catcode `\_12\catcode `\%12\relax}%
\providecommand \@@startlink[1]{}%
\providecommand \@@endlink[0]{}%
\providecommand \url  [0]{\begingroup\@sanitize@url \@url }%
\providecommand \@url [1]{\endgroup\@href {#1}{\urlprefix }}%
\providecommand \urlprefix  [0]{URL }%
\providecommand \Eprint [0]{\href }%
\providecommand \doibase [0]{https://doi.org/}%
\providecommand \selectlanguage [0]{\@gobble}%
\providecommand \bibinfo  [0]{\@secondoftwo}%
\providecommand \bibfield  [0]{\@secondoftwo}%
\providecommand \translation [1]{[#1]}%
\providecommand \BibitemOpen [0]{}%
\providecommand \bibitemStop [0]{}%
\providecommand \bibitemNoStop [0]{.\EOS\space}%
\providecommand \EOS [0]{\spacefactor3000\relax}%
\providecommand \BibitemShut  [1]{\csname bibitem#1\endcsname}%
\let\auto@bib@innerbib\@empty
\bibitem [{\citenamefont {Caloz}\ and\ \citenamefont
  {Deck-Leger}(2019)}]{Caloz2019spacetime}%
  \BibitemOpen
  \bibfield  {author} {\bibinfo {author} {\bibfnamefont {C.}~\bibnamefont
  {Caloz}}\ and\ \bibinfo {author} {\bibfnamefont {Z.-L.}\ \bibnamefont
  {Deck-Leger}},\ }\bibfield  {title} {\bibinfo {title} {Spacetime
  metamaterials—part {II}: theory and applications},\ }\href@noop {}
  {\bibfield  {journal} {\bibinfo  {journal} {IEEE Transactions on Antennas and
  Propagation}\ }\textbf {\bibinfo {volume} {68}},\ \bibinfo {pages} {1583}
  (\bibinfo {year} {2019})}\BibitemShut {NoStop}%
\bibitem [{\citenamefont {Galiffi}\ \emph {et~al.}(2022)\citenamefont
  {Galiffi}, \citenamefont {Tirole}, \citenamefont {Yin}, \citenamefont {Li},
  \citenamefont {Vezzoli}, \citenamefont {Huidobro}, \citenamefont
  {Silveirinha}, \citenamefont {Sapienza}, \citenamefont {Al{\`u}},\ and\
  \citenamefont {Pendry}}]{galiffi2022photonics}%
  \BibitemOpen
  \bibfield  {author} {\bibinfo {author} {\bibfnamefont {E.}~\bibnamefont
  {Galiffi}}, \bibinfo {author} {\bibfnamefont {R.}~\bibnamefont {Tirole}},
  \bibinfo {author} {\bibfnamefont {S.}~\bibnamefont {Yin}}, \bibinfo {author}
  {\bibfnamefont {H.}~\bibnamefont {Li}}, \bibinfo {author} {\bibfnamefont
  {S.}~\bibnamefont {Vezzoli}}, \bibinfo {author} {\bibfnamefont {P.~A.}\
  \bibnamefont {Huidobro}}, \bibinfo {author} {\bibfnamefont {M.~G.}\
  \bibnamefont {Silveirinha}}, \bibinfo {author} {\bibfnamefont
  {R.}~\bibnamefont {Sapienza}}, \bibinfo {author} {\bibfnamefont
  {A.}~\bibnamefont {Al{\`u}}},\ and\ \bibinfo {author} {\bibfnamefont
  {J.}~\bibnamefont {Pendry}},\ }\bibfield  {title} {\bibinfo {title}
  {Photonics of time-varying media},\ }\href@noop {} {\bibfield  {journal}
  {\bibinfo  {journal} {Advanced Photonics}\ }\textbf {\bibinfo {volume} {4}},\
  \bibinfo {pages} {014002} (\bibinfo {year} {2022})}\BibitemShut {NoStop}%
\bibitem [{\citenamefont {Yin}\ \emph {et~al.}(2022)\citenamefont {Yin},
  \citenamefont {Galiffi},\ and\ \citenamefont {Al{\`u}}}]{Yin2022}%
  \BibitemOpen
  \bibfield  {author} {\bibinfo {author} {\bibfnamefont {S.}~\bibnamefont
  {Yin}}, \bibinfo {author} {\bibfnamefont {E.}~\bibnamefont {Galiffi}},\ and\
  \bibinfo {author} {\bibfnamefont {A.}~\bibnamefont {Al{\`u}}},\ }\bibfield
  {title} {\bibinfo {title} {Floquet metamaterials},\ }\href@noop {} {\bibfield
   {journal} {\bibinfo  {journal} {eLight}\ }\textbf {\bibinfo {volume} {2}},\
  \bibinfo {pages} {1} (\bibinfo {year} {2022})}\BibitemShut {NoStop}%
\bibitem [{\citenamefont {Morgenthaler}(1958)}]{Morgenthaler1958velocity}%
  \BibitemOpen
  \bibfield  {author} {\bibinfo {author} {\bibfnamefont {F.~R.}\ \bibnamefont
  {Morgenthaler}},\ }\bibfield  {title} {\bibinfo {title} {Velocity modulation
  of electromagnetic waves},\ }\href@noop {} {\bibfield  {journal} {\bibinfo
  {journal} {IRE Transactions on Microwave Theory and Techniques}\ }\textbf
  {\bibinfo {volume} {6}},\ \bibinfo {pages} {167} (\bibinfo {year}
  {1958})}\BibitemShut {NoStop}%
\bibitem [{\citenamefont {Felsen}\ and\ \citenamefont
  {Whitman}(1970)}]{Felsen1970wave}%
  \BibitemOpen
  \bibfield  {author} {\bibinfo {author} {\bibfnamefont {L.}~\bibnamefont
  {Felsen}}\ and\ \bibinfo {author} {\bibfnamefont {G.}~\bibnamefont
  {Whitman}},\ }\bibfield  {title} {\bibinfo {title} {Wave propagation in
  time-varying media},\ }\href@noop {} {\bibfield  {journal} {\bibinfo
  {journal} {IEEE Transactions on Antennas and Propagation}\ }\textbf {\bibinfo
  {volume} {18}},\ \bibinfo {pages} {242} (\bibinfo {year} {1970})}\BibitemShut
  {NoStop}%
\bibitem [{\citenamefont {Fante}(1971)}]{Fante1971transmission}%
  \BibitemOpen
  \bibfield  {author} {\bibinfo {author} {\bibfnamefont {R.}~\bibnamefont
  {Fante}},\ }\bibfield  {title} {\bibinfo {title} {Transmission of
  electromagnetic waves into time-varying media},\ }\href@noop {} {\bibfield
  {journal} {\bibinfo  {journal} {IEEE Transactions on Antennas and
  Propagation}\ }\textbf {\bibinfo {volume} {19}},\ \bibinfo {pages} {417}
  (\bibinfo {year} {1971})}\BibitemShut {NoStop}%
\bibitem [{\citenamefont {Fante}(1973)}]{Fante1973propagation}%
  \BibitemOpen
  \bibfield  {author} {\bibinfo {author} {\bibfnamefont {R.}~\bibnamefont
  {Fante}},\ }\bibfield  {title} {\bibinfo {title} {On the propagation of
  electromagnetic waves through a time-varying dielectric layer},\ }\href@noop
  {} {\bibfield  {journal} {\bibinfo  {journal} {Applied Scientific Research}\
  }\textbf {\bibinfo {volume} {27}},\ \bibinfo {pages} {341} (\bibinfo {year}
  {1973})}\BibitemShut {NoStop}%
\bibitem [{\citenamefont {Xiao}\ \emph {et~al.}(2014)\citenamefont {Xiao},
  \citenamefont {Maywar},\ and\ \citenamefont {Agrawal}}]{xiao2014reflection}%
  \BibitemOpen
  \bibfield  {author} {\bibinfo {author} {\bibfnamefont {Y.}~\bibnamefont
  {Xiao}}, \bibinfo {author} {\bibfnamefont {D.~N.}\ \bibnamefont {Maywar}},\
  and\ \bibinfo {author} {\bibfnamefont {G.~P.}\ \bibnamefont {Agrawal}},\
  }\bibfield  {title} {\bibinfo {title} {Reflection and transmission of
  electromagnetic waves at a temporal boundary},\ }\href@noop {} {\bibfield
  {journal} {\bibinfo  {journal} {Optics Letters}\ }\textbf {\bibinfo {volume}
  {39}},\ \bibinfo {pages} {574} (\bibinfo {year} {2014})}\BibitemShut
  {NoStop}%
\bibitem [{\citenamefont {Akbarzadeh}\ \emph {et~al.}(2018)\citenamefont
  {Akbarzadeh}, \citenamefont {Chamanara},\ and\ \citenamefont
  {Caloz}}]{akbarzadeh2018inverse}%
  \BibitemOpen
  \bibfield  {author} {\bibinfo {author} {\bibfnamefont {A.}~\bibnamefont
  {Akbarzadeh}}, \bibinfo {author} {\bibfnamefont {N.}~\bibnamefont
  {Chamanara}},\ and\ \bibinfo {author} {\bibfnamefont {C.}~\bibnamefont
  {Caloz}},\ }\bibfield  {title} {\bibinfo {title} {Inverse prism based on
  temporal discontinuity and spatial dispersion},\ }\href@noop {} {\bibfield
  {journal} {\bibinfo  {journal} {Optics Letters}\ }\textbf {\bibinfo {volume}
  {43}},\ \bibinfo {pages} {3297} (\bibinfo {year} {2018})}\BibitemShut
  {NoStop}%
\bibitem [{\citenamefont {Pacheco-Pe{\~n}a}\ and\ \citenamefont
  {Engheta}(2020{\natexlab{a}})}]{Pacheco2020temporal}%
  \BibitemOpen
  \bibfield  {author} {\bibinfo {author} {\bibfnamefont {V.}~\bibnamefont
  {Pacheco-Pe{\~n}a}}\ and\ \bibinfo {author} {\bibfnamefont {N.}~\bibnamefont
  {Engheta}},\ }\bibfield  {title} {\bibinfo {title} {Temporal aiming},\
  }\href@noop {} {\bibfield  {journal} {\bibinfo  {journal} {Light: Science \&
  Applications}\ }\textbf {\bibinfo {volume} {9}},\ \bibinfo {pages} {1}
  (\bibinfo {year} {2020}{\natexlab{a}})}\BibitemShut {NoStop}%
\bibitem [{\citenamefont {Pacheco-Pe{\~n}a}\ and\ \citenamefont
  {Engheta}(2021)}]{pacheco2021Brewster}%
  \BibitemOpen
  \bibfield  {author} {\bibinfo {author} {\bibfnamefont {V.}~\bibnamefont
  {Pacheco-Pe{\~n}a}}\ and\ \bibinfo {author} {\bibfnamefont {N.}~\bibnamefont
  {Engheta}},\ }\bibfield  {title} {\bibinfo {title} {Temporal equivalent of
  the brewster angle},\ }\href@noop {} {\bibfield  {journal} {\bibinfo
  {journal} {Physical Review B}\ }\textbf {\bibinfo {volume} {104}},\ \bibinfo
  {pages} {214308} (\bibinfo {year} {2021})}\BibitemShut {NoStop}%
\bibitem [{\citenamefont {Sol{\'\i}s}\ \emph {et~al.}(2021)\citenamefont
  {Sol{\'\i}s}, \citenamefont {Kastner},\ and\ \citenamefont
  {Engheta}}]{Solis2021time}%
  \BibitemOpen
  \bibfield  {author} {\bibinfo {author} {\bibfnamefont {D.~M.}\ \bibnamefont
  {Sol{\'\i}s}}, \bibinfo {author} {\bibfnamefont {R.}~\bibnamefont
  {Kastner}},\ and\ \bibinfo {author} {\bibfnamefont {N.}~\bibnamefont
  {Engheta}},\ }\bibfield  {title} {\bibinfo {title} {Time-varying materials in
  the presence of dispersion: plane-wave propagation in a lorentzian medium
  with temporal discontinuity},\ }\href@noop {} {\bibfield  {journal} {\bibinfo
   {journal} {Photonics Research}\ }\textbf {\bibinfo {volume} {9}},\ \bibinfo
  {pages} {1842} (\bibinfo {year} {2021})}\BibitemShut {NoStop}%
\bibitem [{\citenamefont {Gratus}\ \emph {et~al.}(2021)\citenamefont {Gratus},
  \citenamefont {Seviour}, \citenamefont {Kinsler},\ and\ \citenamefont
  {Jaroszynski}}]{Gratus2021temporal}%
  \BibitemOpen
  \bibfield  {author} {\bibinfo {author} {\bibfnamefont {J.}~\bibnamefont
  {Gratus}}, \bibinfo {author} {\bibfnamefont {R.}~\bibnamefont {Seviour}},
  \bibinfo {author} {\bibfnamefont {P.}~\bibnamefont {Kinsler}},\ and\ \bibinfo
  {author} {\bibfnamefont {D.~A.}\ \bibnamefont {Jaroszynski}},\ }\bibfield
  {title} {\bibinfo {title} {Temporal boundaries in electromagnetic
  materials},\ }\href@noop {} {\bibfield  {journal} {\bibinfo  {journal} {New
  Journal of Physics}\ }\textbf {\bibinfo {volume} {23}},\ \bibinfo {pages}
  {083032} (\bibinfo {year} {2021})}\BibitemShut {NoStop}%
\bibitem [{\citenamefont {Li}\ \emph {et~al.}(2022)\citenamefont {Li},
  \citenamefont {Yin},\ and\ \citenamefont {Al{\`u}}}]{Li2022nonreciprocity}%
  \BibitemOpen
  \bibfield  {author} {\bibinfo {author} {\bibfnamefont {H.}~\bibnamefont
  {Li}}, \bibinfo {author} {\bibfnamefont {S.}~\bibnamefont {Yin}},\ and\
  \bibinfo {author} {\bibfnamefont {A.}~\bibnamefont {Al{\`u}}},\ }\bibfield
  {title} {\bibinfo {title} {Nonreciprocity and {F}araday rotation at time
  interfaces},\ }\href@noop {} {\bibfield  {journal} {\bibinfo  {journal}
  {Physical Review Letters}\ }\textbf {\bibinfo {volume} {128}},\ \bibinfo
  {pages} {173901} (\bibinfo {year} {2022})}\BibitemShut {NoStop}%
\bibitem [{\citenamefont {Pacheco-Pe{\~n}a}\ and\ \citenamefont
  {Engheta}(2020{\natexlab{b}})}]{Pacheco2020antireflection}%
  \BibitemOpen
  \bibfield  {author} {\bibinfo {author} {\bibfnamefont {V.}~\bibnamefont
  {Pacheco-Pe{\~n}a}}\ and\ \bibinfo {author} {\bibfnamefont {N.}~\bibnamefont
  {Engheta}},\ }\bibfield  {title} {\bibinfo {title} {Antireflection temporal
  coatings},\ }\href@noop {} {\bibfield  {journal} {\bibinfo  {journal}
  {Optica}\ }\textbf {\bibinfo {volume} {7}},\ \bibinfo {pages} {323} (\bibinfo
  {year} {2020}{\natexlab{b}})}\BibitemShut {NoStop}%
\bibitem [{\citenamefont {Castaldi}\ \emph {et~al.}(2021)\citenamefont
  {Castaldi}, \citenamefont {Pacheco-Pe{\~n}a}, \citenamefont {Moccia},
  \citenamefont {Engheta},\ and\ \citenamefont
  {Galdi}}]{Castaldi2021exploiting}%
  \BibitemOpen
  \bibfield  {author} {\bibinfo {author} {\bibfnamefont {G.}~\bibnamefont
  {Castaldi}}, \bibinfo {author} {\bibfnamefont {V.}~\bibnamefont
  {Pacheco-Pe{\~n}a}}, \bibinfo {author} {\bibfnamefont {M.}~\bibnamefont
  {Moccia}}, \bibinfo {author} {\bibfnamefont {N.}~\bibnamefont {Engheta}},\
  and\ \bibinfo {author} {\bibfnamefont {V.}~\bibnamefont {Galdi}},\ }\bibfield
   {title} {\bibinfo {title} {Exploiting space-time duality in the synthesis of
  impedance transformers via temporal metamaterials},\ }\href@noop {}
  {\bibfield  {journal} {\bibinfo  {journal} {Nanophotonics}\ }\textbf
  {\bibinfo {volume} {10}},\ \bibinfo {pages} {3687} (\bibinfo {year}
  {2021})}\BibitemShut {NoStop}%
\bibitem [{\citenamefont {Ramaccia}\ \emph {et~al.}(2020)\citenamefont
  {Ramaccia}, \citenamefont {Toscano},\ and\ \citenamefont
  {Bilotti}}]{Ramaccia2020light}%
  \BibitemOpen
  \bibfield  {author} {\bibinfo {author} {\bibfnamefont {D.}~\bibnamefont
  {Ramaccia}}, \bibinfo {author} {\bibfnamefont {A.}~\bibnamefont {Toscano}},\
  and\ \bibinfo {author} {\bibfnamefont {F.}~\bibnamefont {Bilotti}},\
  }\bibfield  {title} {\bibinfo {title} {Light propagation through metamaterial
  temporal slabs: Reflection, refraction, and special cases},\ }\href@noop {}
  {\bibfield  {journal} {\bibinfo  {journal} {Optics Letters}\ }\textbf
  {\bibinfo {volume} {45}},\ \bibinfo {pages} {5836} (\bibinfo {year}
  {2020})}\BibitemShut {NoStop}%
\bibitem [{\citenamefont {Ramaccia}\ \emph {et~al.}(2021)\citenamefont
  {Ramaccia}, \citenamefont {Al{\`u}}, \citenamefont {Toscano},\ and\
  \citenamefont {Bilotti}}]{Ramaccia2021temporal}%
  \BibitemOpen
  \bibfield  {author} {\bibinfo {author} {\bibfnamefont {D.}~\bibnamefont
  {Ramaccia}}, \bibinfo {author} {\bibfnamefont {A.}~\bibnamefont {Al{\`u}}},
  \bibinfo {author} {\bibfnamefont {A.}~\bibnamefont {Toscano}},\ and\ \bibinfo
  {author} {\bibfnamefont {F.}~\bibnamefont {Bilotti}},\ }\bibfield  {title}
  {\bibinfo {title} {Temporal multilayer structures for designing higher-order
  transfer functions using time-varying metamaterials},\ }\href@noop {}
  {\bibfield  {journal} {\bibinfo  {journal} {Applied Physics Letters}\
  }\textbf {\bibinfo {volume} {118}},\ \bibinfo {pages} {101901} (\bibinfo
  {year} {2021})}\BibitemShut {NoStop}%
\bibitem [{\citenamefont {Lustig}\ \emph {et~al.}(2018)\citenamefont {Lustig},
  \citenamefont {Sharabi},\ and\ \citenamefont
  {Segev}}]{Lustig2018topological}%
  \BibitemOpen
  \bibfield  {author} {\bibinfo {author} {\bibfnamefont {E.}~\bibnamefont
  {Lustig}}, \bibinfo {author} {\bibfnamefont {Y.}~\bibnamefont {Sharabi}},\
  and\ \bibinfo {author} {\bibfnamefont {M.}~\bibnamefont {Segev}},\ }\bibfield
   {title} {\bibinfo {title} {Topological aspects of photonic time crystals},\
  }\href@noop {} {\bibfield  {journal} {\bibinfo  {journal} {Optica}\ }\textbf
  {\bibinfo {volume} {5}},\ \bibinfo {pages} {1390} (\bibinfo {year}
  {2018})}\BibitemShut {NoStop}%
\bibitem [{\citenamefont {Sharabi}\ \emph {et~al.}(2021)\citenamefont
  {Sharabi}, \citenamefont {Lustig},\ and\ \citenamefont
  {Segev}}]{Sharabi2021disordered}%
  \BibitemOpen
  \bibfield  {author} {\bibinfo {author} {\bibfnamefont {Y.}~\bibnamefont
  {Sharabi}}, \bibinfo {author} {\bibfnamefont {E.}~\bibnamefont {Lustig}},\
  and\ \bibinfo {author} {\bibfnamefont {M.}~\bibnamefont {Segev}},\ }\bibfield
   {title} {\bibinfo {title} {Disordered photonic time crystals},\ }\href@noop
  {} {\bibfield  {journal} {\bibinfo  {journal} {Physical Review Letters}\
  }\textbf {\bibinfo {volume} {126}},\ \bibinfo {pages} {163902} (\bibinfo
  {year} {2021})}\BibitemShut {NoStop}%
\bibitem [{\citenamefont {Sharabi}\ \emph {et~al.}(2022)\citenamefont
  {Sharabi}, \citenamefont {Dikopoltsev}, \citenamefont {Lustig}, \citenamefont
  {Lumer},\ and\ \citenamefont {Segev}}]{Sharabi2022spatiotemporal}%
  \BibitemOpen
  \bibfield  {author} {\bibinfo {author} {\bibfnamefont {Y.}~\bibnamefont
  {Sharabi}}, \bibinfo {author} {\bibfnamefont {A.}~\bibnamefont
  {Dikopoltsev}}, \bibinfo {author} {\bibfnamefont {E.}~\bibnamefont {Lustig}},
  \bibinfo {author} {\bibfnamefont {Y.}~\bibnamefont {Lumer}},\ and\ \bibinfo
  {author} {\bibfnamefont {M.}~\bibnamefont {Segev}},\ }\bibfield  {title}
  {\bibinfo {title} {Spatiotemporal photonic crystals},\ }\href@noop {}
  {\bibfield  {journal} {\bibinfo  {journal} {Optica}\ }\textbf {\bibinfo
  {volume} {9}},\ \bibinfo {pages} {585} (\bibinfo {year} {2022})}\BibitemShut
  {NoStop}%
\bibitem [{\citenamefont {Huidobro}\ \emph {et~al.}(2021)\citenamefont
  {Huidobro}, \citenamefont {Silveirinha}, \citenamefont {Galiffi},\ and\
  \citenamefont {Pendry}}]{Huidobro2021homogenization}%
  \BibitemOpen
  \bibfield  {author} {\bibinfo {author} {\bibfnamefont {P.~A.}\ \bibnamefont
  {Huidobro}}, \bibinfo {author} {\bibfnamefont {M.~G.}\ \bibnamefont
  {Silveirinha}}, \bibinfo {author} {\bibfnamefont {E.}~\bibnamefont
  {Galiffi}},\ and\ \bibinfo {author} {\bibfnamefont {J.}~\bibnamefont
  {Pendry}},\ }\bibfield  {title} {\bibinfo {title} {Homogenization theory of
  space-time metamaterials},\ }\href@noop {} {\bibfield  {journal} {\bibinfo
  {journal} {Physical Review Applied}\ }\textbf {\bibinfo {volume} {16}},\
  \bibinfo {pages} {014044} (\bibinfo {year} {2021})}\BibitemShut {NoStop}%
\bibitem [{\citenamefont {Mendon{\c{c}}a}\ \emph {et~al.}(2000)\citenamefont
  {Mendon{\c{c}}a}, \citenamefont {Guerreiro},\ and\ \citenamefont
  {Martins}}]{Mendoncca2000quantum}%
  \BibitemOpen
  \bibfield  {author} {\bibinfo {author} {\bibfnamefont {J.}~\bibnamefont
  {Mendon{\c{c}}a}}, \bibinfo {author} {\bibfnamefont {A.}~\bibnamefont
  {Guerreiro}},\ and\ \bibinfo {author} {\bibfnamefont {A.~M.}\ \bibnamefont
  {Martins}},\ }\bibfield  {title} {\bibinfo {title} {Quantum theory of time
  refraction},\ }\href@noop {} {\bibfield  {journal} {\bibinfo  {journal}
  {Physical Review A}\ }\textbf {\bibinfo {volume} {62}},\ \bibinfo {pages}
  {033805} (\bibinfo {year} {2000})}\BibitemShut {NoStop}%
\bibitem [{\citenamefont {Mendon{\c{c}}a}\ \emph {et~al.}(2003)\citenamefont
  {Mendon{\c{c}}a}, \citenamefont {Martins},\ and\ \citenamefont
  {Guerreiro}}]{Mendoncca2003temporal}%
  \BibitemOpen
  \bibfield  {author} {\bibinfo {author} {\bibfnamefont {J.}~\bibnamefont
  {Mendon{\c{c}}a}}, \bibinfo {author} {\bibfnamefont {A.}~\bibnamefont
  {Martins}},\ and\ \bibinfo {author} {\bibfnamefont {A.}~\bibnamefont
  {Guerreiro}},\ }\bibfield  {title} {\bibinfo {title} {Temporal beam splitter
  and temporal interference},\ }\href@noop {} {\bibfield  {journal} {\bibinfo
  {journal} {Physical Review A}\ }\textbf {\bibinfo {volume} {68}},\ \bibinfo
  {pages} {043801} (\bibinfo {year} {2003})}\BibitemShut {NoStop}%
\bibitem [{\citenamefont {Mendon{\c{c}}a}\ and\ \citenamefont
  {Guerreiro}(2005)}]{Mendoncca2005time}%
  \BibitemOpen
  \bibfield  {author} {\bibinfo {author} {\bibfnamefont {J.}~\bibnamefont
  {Mendon{\c{c}}a}}\ and\ \bibinfo {author} {\bibfnamefont {A.}~\bibnamefont
  {Guerreiro}},\ }\bibfield  {title} {\bibinfo {title} {Time refraction and the
  quantum properties of vacuum},\ }\href@noop {} {\bibfield  {journal}
  {\bibinfo  {journal} {Physical Review A}\ }\textbf {\bibinfo {volume} {72}},\
  \bibinfo {pages} {063805} (\bibinfo {year} {2005})}\BibitemShut {NoStop}%
\bibitem [{\citenamefont {Lyubarov}\ \emph {et~al.}(2022)\citenamefont
  {Lyubarov}, \citenamefont {Lumer}, \citenamefont {Dikopoltsev}, \citenamefont
  {Lustig}, \citenamefont {Sharabi},\ and\ \citenamefont
  {Segev}}]{Lyubarov2022amplified}%
  \BibitemOpen
  \bibfield  {author} {\bibinfo {author} {\bibfnamefont {M.}~\bibnamefont
  {Lyubarov}}, \bibinfo {author} {\bibfnamefont {Y.}~\bibnamefont {Lumer}},
  \bibinfo {author} {\bibfnamefont {A.}~\bibnamefont {Dikopoltsev}}, \bibinfo
  {author} {\bibfnamefont {E.}~\bibnamefont {Lustig}}, \bibinfo {author}
  {\bibfnamefont {Y.}~\bibnamefont {Sharabi}},\ and\ \bibinfo {author}
  {\bibfnamefont {M.}~\bibnamefont {Segev}},\ }\bibfield  {title} {\bibinfo
  {title} {Amplified emission by atoms and lasing in photonic time crystals},\
  }\href@noop {} {\bibfield  {journal} {\bibinfo  {journal} {arXiv preprint
  arXiv:2201.12116}\ } (\bibinfo {year} {2022})}\BibitemShut {NoStop}%
\bibitem [{\citenamefont {Dikopoltsev}\ \emph {et~al.}(2022)\citenamefont
  {Dikopoltsev}, \citenamefont {Sharabi}, \citenamefont {Lyubarov},
  \citenamefont {Lumer}, \citenamefont {Tsesses}, \citenamefont {Lustig},
  \citenamefont {Kaminer},\ and\ \citenamefont {Segev}}]{Dikopoltsev2022light}%
  \BibitemOpen
  \bibfield  {author} {\bibinfo {author} {\bibfnamefont {A.}~\bibnamefont
  {Dikopoltsev}}, \bibinfo {author} {\bibfnamefont {Y.}~\bibnamefont
  {Sharabi}}, \bibinfo {author} {\bibfnamefont {M.}~\bibnamefont {Lyubarov}},
  \bibinfo {author} {\bibfnamefont {Y.}~\bibnamefont {Lumer}}, \bibinfo
  {author} {\bibfnamefont {S.}~\bibnamefont {Tsesses}}, \bibinfo {author}
  {\bibfnamefont {E.}~\bibnamefont {Lustig}}, \bibinfo {author} {\bibfnamefont
  {I.}~\bibnamefont {Kaminer}},\ and\ \bibinfo {author} {\bibfnamefont
  {M.}~\bibnamefont {Segev}},\ }\bibfield  {title} {\bibinfo {title} {Light
  emission by free electrons in photonic time-crystals},\ }\href@noop {}
  {\bibfield  {journal} {\bibinfo  {journal} {Proceedings of the National
  Academy of Sciences}\ }\textbf {\bibinfo {volume} {119}},\ \bibinfo {pages}
  {e2119705119} (\bibinfo {year} {2022})}\BibitemShut {NoStop}%
\bibitem [{\citenamefont {Nation}\ \emph {et~al.}(2012)\citenamefont {Nation},
  \citenamefont {Johansson}, \citenamefont {Blencowe},\ and\ \citenamefont
  {Nori}}]{Nation2012colloquium}%
  \BibitemOpen
  \bibfield  {author} {\bibinfo {author} {\bibfnamefont {P.}~\bibnamefont
  {Nation}}, \bibinfo {author} {\bibfnamefont {J.}~\bibnamefont {Johansson}},
  \bibinfo {author} {\bibfnamefont {M.}~\bibnamefont {Blencowe}},\ and\
  \bibinfo {author} {\bibfnamefont {F.}~\bibnamefont {Nori}},\ }\bibfield
  {title} {\bibinfo {title} {Colloquium: Stimulating uncertainty: Amplifying
  the quantum vacuum with superconducting circuits},\ }\href@noop {} {\bibfield
   {journal} {\bibinfo  {journal} {Reviews of Modern Physics}\ }\textbf
  {\bibinfo {volume} {84}},\ \bibinfo {pages} {1} (\bibinfo {year}
  {2012})}\BibitemShut {NoStop}%
\bibitem [{\citenamefont {Dodonov}(2020)}]{Dodonov2020fifty}%
  \BibitemOpen
  \bibfield  {author} {\bibinfo {author} {\bibfnamefont {V.}~\bibnamefont
  {Dodonov}},\ }\bibfield  {title} {\bibinfo {title} {Fifty years of the
  dynamical {C}asimir effect},\ }\href@noop {} {\bibfield  {journal} {\bibinfo
  {journal} {Physics}\ }\textbf {\bibinfo {volume} {2}},\ \bibinfo {pages} {67}
  (\bibinfo {year} {2020})}\BibitemShut {NoStop}%
\bibitem [{\citenamefont {Wilson}\ \emph {et~al.}(2011)\citenamefont {Wilson},
  \citenamefont {Johansson}, \citenamefont {Pourkabirian}, \citenamefont
  {Simoen}, \citenamefont {Johansson}, \citenamefont {Duty}, \citenamefont
  {Nori},\ and\ \citenamefont {Delsing}}]{Wilson2011observation}%
  \BibitemOpen
  \bibfield  {author} {\bibinfo {author} {\bibfnamefont {C.~M.}\ \bibnamefont
  {Wilson}}, \bibinfo {author} {\bibfnamefont {G.}~\bibnamefont {Johansson}},
  \bibinfo {author} {\bibfnamefont {A.}~\bibnamefont {Pourkabirian}}, \bibinfo
  {author} {\bibfnamefont {M.}~\bibnamefont {Simoen}}, \bibinfo {author}
  {\bibfnamefont {J.~R.}\ \bibnamefont {Johansson}}, \bibinfo {author}
  {\bibfnamefont {T.}~\bibnamefont {Duty}}, \bibinfo {author} {\bibfnamefont
  {F.}~\bibnamefont {Nori}},\ and\ \bibinfo {author} {\bibfnamefont
  {P.}~\bibnamefont {Delsing}},\ }\bibfield  {title} {\bibinfo {title}
  {Observation of the dynamical {C}asimir effect in a superconducting
  circuit},\ }\href@noop {} {\bibfield  {journal} {\bibinfo  {journal}
  {Nature}\ }\textbf {\bibinfo {volume} {479}},\ \bibinfo {pages} {376}
  (\bibinfo {year} {2011})}\BibitemShut {NoStop}%
\bibitem [{\citenamefont {L{\"a}hteenm{\"a}ki}\ \emph
  {et~al.}(2013)\citenamefont {L{\"a}hteenm{\"a}ki}, \citenamefont {Paraoanu},
  \citenamefont {Hassel},\ and\ \citenamefont
  {Hakonen}}]{Lahteenmaki2013dynamical}%
  \BibitemOpen
  \bibfield  {author} {\bibinfo {author} {\bibfnamefont {P.}~\bibnamefont
  {L{\"a}hteenm{\"a}ki}}, \bibinfo {author} {\bibfnamefont {G.}~\bibnamefont
  {Paraoanu}}, \bibinfo {author} {\bibfnamefont {J.}~\bibnamefont {Hassel}},\
  and\ \bibinfo {author} {\bibfnamefont {P.~J.}\ \bibnamefont {Hakonen}},\
  }\bibfield  {title} {\bibinfo {title} {Dynamical {C}asimir effect in a
  {J}osephson metamaterial},\ }\href@noop {} {\bibfield  {journal} {\bibinfo
  {journal} {Proceedings of the National Academy of Sciences}\ }\textbf
  {\bibinfo {volume} {110}},\ \bibinfo {pages} {4234} (\bibinfo {year}
  {2013})}\BibitemShut {NoStop}%
\bibitem [{\citenamefont {Caloz}\ and\ \citenamefont {Itoh}(2005)}]{Caloz2005}%
  \BibitemOpen
  \bibfield  {author} {\bibinfo {author} {\bibfnamefont {C.}~\bibnamefont
  {Caloz}}\ and\ \bibinfo {author} {\bibfnamefont {T.}~\bibnamefont {Itoh}},\
  }\href@noop {} {\emph {\bibinfo {title} {Electromagnetic metamaterials:
  transmission line theory and microwave applications}}}\ (\bibinfo
  {publisher} {John Wiley \& Sons},\ \bibinfo {year} {2005})\BibitemShut
  {NoStop}%
\bibitem [{\citenamefont {Eleftheriades}\ and\ \citenamefont
  {Balmain}(2005)}]{Eleftheriades2005negative}%
  \BibitemOpen
  \bibfield  {author} {\bibinfo {author} {\bibfnamefont {G.~V.}\ \bibnamefont
  {Eleftheriades}}\ and\ \bibinfo {author} {\bibfnamefont {K.~G.}\ \bibnamefont
  {Balmain}},\ }\href@noop {} {\emph {\bibinfo {title} {Negative-refraction
  metamaterials: fundamental principles and applications}}}\ (\bibinfo
  {publisher} {John Wiley \& Sons},\ \bibinfo {year} {2005})\BibitemShut
  {NoStop}%
\bibitem [{\citenamefont {Grbic}\ and\ \citenamefont
  {Eleftheriades}(2004)}]{Grbic2004overcoming}%
  \BibitemOpen
  \bibfield  {author} {\bibinfo {author} {\bibfnamefont {A.}~\bibnamefont
  {Grbic}}\ and\ \bibinfo {author} {\bibfnamefont {G.~V.}\ \bibnamefont
  {Eleftheriades}},\ }\bibfield  {title} {\bibinfo {title} {Overcoming the
  diffraction limit with a planar left-handed transmission-line lens},\
  }\href@noop {} {\bibfield  {journal} {\bibinfo  {journal} {Physical review
  letters}\ }\textbf {\bibinfo {volume} {92}},\ \bibinfo {pages} {117403}
  (\bibinfo {year} {2004})}\BibitemShut {NoStop}%
\bibitem [{\citenamefont {Zedler}\ and\ \citenamefont
  {Eleftheriades}(2011)}]{Zedler2011anisotropic}%
  \BibitemOpen
  \bibfield  {author} {\bibinfo {author} {\bibfnamefont {M.}~\bibnamefont
  {Zedler}}\ and\ \bibinfo {author} {\bibfnamefont {G.~V.}\ \bibnamefont
  {Eleftheriades}},\ }\bibfield  {title} {\bibinfo {title} {Anisotropic
  transmission-line metamaterials for 2-d transformation optics applications},\
  }\href@noop {} {\bibfield  {journal} {\bibinfo  {journal} {Proceedings of the
  IEEE}\ }\textbf {\bibinfo {volume} {99}},\ \bibinfo {pages} {1634} (\bibinfo
  {year} {2011})}\BibitemShut {NoStop}%
\bibitem [{\citenamefont {Wong}\ \emph {et~al.}(2006)\citenamefont {Wong},
  \citenamefont {Balmain},\ and\ \citenamefont
  {Eleftheriades}}]{Wong2006fields}%
  \BibitemOpen
  \bibfield  {author} {\bibinfo {author} {\bibfnamefont {J.~K.}\ \bibnamefont
  {Wong}}, \bibinfo {author} {\bibfnamefont {K.~G.}\ \bibnamefont {Balmain}},\
  and\ \bibinfo {author} {\bibfnamefont {G.~V.}\ \bibnamefont
  {Eleftheriades}},\ }\bibfield  {title} {\bibinfo {title} {Fields in planar
  anisotropic transmission-line metamaterials},\ }\href@noop {} {\bibfield
  {journal} {\bibinfo  {journal} {IEEE Transactions on Antennas and
  Propagation}\ }\textbf {\bibinfo {volume} {54}},\ \bibinfo {pages} {2742}
  (\bibinfo {year} {2006})}\BibitemShut {NoStop}%
\bibitem [{\citenamefont {Sakurai}\ and\ \citenamefont
  {Napolitano}(2017)}]{Sakurai2014modern}%
  \BibitemOpen
  \bibfield  {author} {\bibinfo {author} {\bibfnamefont {J.~J.~S.}\
  \bibnamefont {Sakurai}}\ and\ \bibinfo {author} {\bibfnamefont {J.~J.}\
  \bibnamefont {Napolitano}},\ }\href@noop {} {\emph {\bibinfo {title} {Modern
  Quantum Mechanics}}}\ (\bibinfo  {publisher} {Cambridge University Press},\
  \bibinfo {year} {2017})\BibitemShut {NoStop}%
\end{thebibliography}%

\end{document}